\documentclass[]{aastex}

\usepackage{emulateapj5}
\usepackage{onecolfloat}
\usepackage{graphicx} 
\usepackage{fancyheadings} 
\usepackage{ulem}
\usepackage{rotating}
\usepackage{lscape}

\newcommand{\ndla}{27}
\newcommand{\kms}{km~s$^{-1}$ }
\newcommand{\cm}[1]{\, {\rm cm^{#1}}}

\newcommand{\dell}{\Delta \lambda}
\newcommand{\delz}{\Delta z}

\newcommand{\lya}{Ly$\alpha$}

\newcommand{\N}[1]{{N({\rm #1})}}

\newcommand{\naj}{AJ.}
\newcommand{\napj}{ApJ.}
\newcommand{\nhi}{$N_{\rm HI}$}
\newcommand{\mnhi}{N_{\rm HI}}

\begin{document}

\twocolumn[%
\submitted{Accepted to the PASP: July 12, 2006}
\title{The Metal-Strong Damped \lya\ Systems}

\author{St\'ephane Herbert-Fort$^{1,2,3}$, 
Jason X. Prochaska$^{2,3}$, 
Miroslava Dessauges-Zavadsky$^{4}$, 
Sara L. Ellison$^{5}$, 
J. Chris Howk$^{6,7}$, 
Arthur M. Wolfe$^{7}$, and
Gabriel E. Prochter$^{2,3}$}
\affil{
$^1$(University of Arizona/Steward Observatory, 933 N Cherry Avenue, Tucson, AZ 85721; shf@as.arizona.edu),
$^2$(University of California Observatories/Lick Observatory, University of California, 1156 High Street, Santa Cruz, CA 95064; xavier@ucolick.org, prochter@ucolick.org), 
$^3$(Visiting Astronomer, W.M. Keck Observatory, a joint facility of the University of California, 
the California Institute of Technology, and NASA), 
$^4$(Observatoire de Gen\`eve, 51 Ch.\ des Maillettes, 1290 Sauverny, Switzerland; Miroslava.Dessauges@obs.unige.ch),
$^5$(University of Victoria, 3800 Finnerty Rd, Victoria, BC V8P 1A1, Canada; sarae@uvic.ca), 
$^6$(Department of Physics, University of Notre Dame, Notre Dame, IN 46556; jhowk@nd.edu), 
$^7$(Department of Physics and Center for Astrophysics and Space Sciences, University 
of California at San Diego, Code 0424, 9500 Gilman Drive, La Jolla, CA 92093; awolfe@ucsd.edu) 
}

\begin{abstract}
We have identified a metal-strong (log$\N{Zn^+} \ge 13.15$ {\it{or}} 
log$\N{Si^+} \ge 15.95$) 
DLA (MSDLA) population from an automated quasar (QSO) absorber search in the 
Sloan Digital Sky Survey Data Release 3 (SDSS-DR3) quasar sample, and 
find that MSDLAs comprise $\approx 5\%$ of the entire DLA population 
with $z_{abs} \ge 2.2$ found in QSO sightlines with $r < 19.5$. 
We have also acquired \ndla\ Keck ESI follow-up spectra of 
metal-strong candidates to evaluate our automated technique 
and examine the MSDLA candidates at higher resolution.
We demonstrate that the rest equivalent widths of 
strong \ion{Zn}{2} $\lambda$2026 and \ion{Si}{2} $\lambda$1808 
lines in low-resolution SDSS spectra 
are accurate 
metal-strong indicators for higher-resolution 
spectra, and predict the observed equivalent widths $W_{obs}$ and 
signal-to-noise ratios (SNRs) needed to detect 
certain extremely weak lines with high-resolution instruments.
We investigate how the MSDLAs may affect 
previous studies concerning a dust-obscuration bias and the $\N{HI}$-weighted 
cosmic mean metallicity $<Z(z)>$.
Finally, we include a brief discussion of abundance ratios in our 
ESI sample and find that 
underlying mostly Type II supernovae enrichment are 
differential depletion effects due to dust (and in a few cases quite strong); 
we present here a handful of 
new Ti and Mn measurements, both of which are 
useful probes of depletion in DLAs.
Future papers will present detailed examinations of particularly 
metal-strong DLAs from high-resolution KeckI/HIRES and VLT/UVES spectra.


\end{abstract}
]

\pagestyle{fancyplain}
\lhead[\fancyplain{}{\thepage}]{\fancyplain{}{HERBERT-FORT et al.}}
\rhead[\fancyplain{}{The Metal-Strong DLAs }]{\fancyplain{}{\thepage}}
\setlength{\headrulewidth=0pt}
\cfoot{}

\section{Introduction}

Damped \lya\ systems (DLAs) are the subset of quasar absorption line (QAL) systems 
classically defined to have neutral hydrogen column densities 
$\mnhi \ge 2 \times 10^{20}$ atoms cm$^{-2}$ \citep{wgp05}.  
They are identified by their wide damped \lya\ absorption profiles, and all DLAs
(to date) show associated metal-line absorption \citep{pro03a}.  
DLAs dominate the neutral gas 
content of the universe and may be expected to constitute the primary 
reservoir of star-forming gas at high redshift 
(Wolfe et al.\ 1995; Prochaska and Herbert-Fort 2004; 
Prochaska, Herbert-Fort and Wolfe 2005, hereafter PHW05).
Therefore, measurements of DLA chemical abundances at high redshift 
help quantify the chemical evolution of the young universe.

Echelle observations of DLAs allow one to accurately measure the
gas-phase abundances of a number of elements and thereby examine 
processes of nucleosynthetic enrichment and differential depletion
in these galaxies \citep[e.g.][]{lu96,vladilo01,pw02,dessauges03}.
However, \cite{pettini94} showed that high-redshift DLAs are generally metal-poor. 
Their results and subsequent studies \citep{pettini99, pro03a, kulkarni05} have tracked the
enrichment of the ISM of galaxies reaching back to the first few Gyr.  
The majority of DLAs show detections of \ion{Fe}{2}, \ion{Ni}{2}, \ion{Si}{2},
and \ion{Al}{2} transitions.  It is unfortunate that the signatures of
Type\,II SNe enrichment \citep{ww95} and differential depletion 
\citep[e.g.][]{savage96} are nearly degenerate
for this small set of elements.  As such, progress in interpreting
the gas-phase abundance patterns of the damped \lya\ systems has been
difficult, although recent works on S, Zn, N and O have made advances.

Prochaska, Howk, and Wolfe (2003) reported the discovery of a
metal-strong DLA at 
$z_{abs}=2.626$ towards the quasar FJ0812+32 (hereafter DLA-B/FJ0812+32).  
In contrast with the majority of damped \lya\ systems, the authors detected
over 20 elements in this single DLA system and revealed the detailed
chemical enrichment pattern of this galaxy.  Many of the detected transitions 
had never before been observed outside of the Local Group and are 
important diagnostics to theories of nucleosynthesis and galaxy enrichment.  
The authors suggested that this system was 
enriched mainly by short-lived, massive stars and 
that it is the progenitor of a 
massive elliptical galaxy \citep[see also][]{fenner04}.  
The goal of this work is to speed the discovery and analysis
of more systems like DLA-B/FJ0812+32; these rare 
DLAs are unique laboratories for the study of
nucleosynthesis, galaxy enrichment, dust depletion, and ISM physics in 
the high-redshift universe.
We hereafter refer to this special subset of DLA as the
metal-strong DLA (MSDLA) systems.


Whilst our primary
motivation for defining MSDLAs as those absorbers with high metal
column densities (see $\S$~2), we note that our definition also corresponds
to an empirical 
upper bound to $\N{Zn^+}$ noted by \cite{boisse98}
for a sample of DLAs in the literature at that time.  Boiss\'e et al.\ 
interpreted the upper bound to $\N{Zn^+}$ as a selection
bias related to dust obscuration (i.e.\ very large dust-to-gas ratio).  
However,
our statistics on MSDLAs (in sightlines with $r < 19.5$) 
now lead us to argue that the paucity of
high N(Zn) absorbers is not due to dust, but simply an indication
of their intrinsic rarity (see also Johansson \& Efstathiou 2006).

We will show that only a few percent of all DLAs are truly metal-strong, and so 
thousands of quasar sightlines must be searched in order to discover just a handful 
of MSDLAs.  Therefore, automated detection algorithms 
used on large quasar surveys are critical to metal-strong DLA research. 
This paper presents our automated method of detecting metal-strong 
absorbers in low-resolution Sloan Digital Sky Survey (SDSS) quasar spectra.  
We detail and release our search algorithms and present 
all of the metal-strong candidates from SDSS Data Release Three
\citep[SDSS-DR3;][]{abazajian05}.

As mentioned above, the elemental abundances from 
DLAs can be used to constrain and test processes of nucleosynthesis. 
For example, 
there are several different theories on the production of Boron.  
\cite{woosley90}
suggested that B production results from 
neutrino spallation in the carbon shells of SNe, while \cite{casse95} have
argued 
for the spallation of C and O nuclei accelerated by SNe onto local interstellar gas.  Other 
theories involve protons and neutrons being accelerated onto interstellar CNO seed nuclei, 
and each theory predicts how B may scale with the galaxy's metallicity (and so here, for 
example, one 
must eventually acquire $\N{HI}$, 
although it is not necessary for the project at hand).  
Measurements of such elements in DLA systems can thus help distinguish between the
various theories. 
Other elements measured in the Galactic ISM would also impact our
understanding of nucleosynthesis and star formation in young galaxies
if these elements were observed in high redshift DLA.
These include: (1) O --  an unambiguous $\alpha$-element 
and the most abundant metal in the universe; (2) Sn and Kr --  r-process 
elements (rapid 
neutron capture in high density and temperature regions);  and 
(3) Pb -- an s-process element 
(slow neutron decay and capture in low density and temperature regions).  
Unfortunately, these elements 
are rarely detected in typical DLA spectra because they have small absolute
abundances and/or their dominant ions have transitions with either too
large or small oscillator strengths 
(e.g.\ \ion{O}{1} $\lambda\lambda 1302,1355$).  In MSDLAs, however, weaker transitions 
become available, allowing high-redshift studies of the processes mentioned 
above.


To gauge the success rate of our algorithms, 
we have obtained moderate-resolution follow-up observations of a subset of the
SDSS-DR3 MSDLA candidates with the 
\\Echelle Spectrograph and 
Imager \citep[][ESI]{sheinis02} on the 10m-class Keck\,II telescope. 
We present our ESI spectra of \ndla\ MSDLA candidates 
($1.6 \le z_{abs} \le 3.1$ and $r < 19.5$) and 
discuss the implications of this candidate metal-strong subsample.

We define the MSDLAs and discuss the impact of \lya\ on our study in $\S$~2, 
present our automated technique for detecting metal-strong 
systems in SDSS, describe how we compiled our sample for medium-resolution 
observations, and report our SDSS search success rates in $\S$~3.
Section 4 gives a summary of our ESI follow-up observations, data reduction, and measurements.  
$\S$~5 presents our analysis and discussion of metal-strong indicators in SDSS QSOs, using our 
ESI sample as a reference.  
Section 6 derives our predictions for detecting certain extremely weak lines with HIRES.  
Section 7 discusses our metal-strong sample and its relation to the proposed dust-obscuration 
bias, as well as a preliminary investigation of how this 
MSDLA population might influence previous detections of an evolution in 
the $\mnhi$-weighted cosmic mean metallicity $<Z(z)>$.
Section 8 concludes the discussion with a few abundance ratios from our ESI data 
with comments on SNe enrichment and depletion due to dust, and 
our summary and conclusions are blended together in $\S$~9.

\section{MSDLA Definition and the Impact of \lya}

With this paper we define the metal-strong 
DLAs (MSDLAs) to have log$\N{Zn^+} \ge 13.15$ 
{\it{or}} log$\N{Si^+} \ge 15.95$, based on the \ion{Zn}{2} 
and \ion{Si}{2} transitions measured in 
\\KeckI/HIRES data of DLA-B/FJ0812+32.
\footnote{Note that the Zn criterion for 
a solar Si/Zn ratio \citep{grvss96} implies 
log$\N{Si^+} \ge 16.0$, and so we see that assuming solar Si/Zn for 
DLA-B/FJ0812+32 is not well supported 
(yet not glaring in difference).  The 
difference may be attributed to depletion effects due to dust, and we 
comment on the role 
of dust in MSDLAs (and in the overall DLA population) in $\S$~7.}
These values are somewhat arbitrary, but chosen because they
imply equivalent widths for weak transitions like \ion{B}{2}~1362 that
can be detected with current 10m-class telescopes.
We have specifically chosen to define the MSDLA subset based on column density 
thresholds and not metallicity (ie.\ not [Zn/H] or [Si/H]), 
due to motivations from 
the DLA-B/FJ0812+32 study.  Specifically, we aim to discover systems which 
may be used as high-redshift probes of the production of elements like 
B, O and Ge (among others) independent of the \nhi\ value.  
Therefore, we do not require having \lya\ measurements and 
corresponding metallicities (ie. for targeting metal-strong systems), although 
having \nhi\ will eventually allow the calculations of ionization fractions 
and dust-to-gas ratios.
We caution that choosing a metal column density threshold 
leads to a mixture of high-\nhi, 
low-metallicity, and low-\nhi, high-metallicity systems which may
be very different in their properties.  This may affect our conclusions 
about the nature or 
evolution of the metal-strong systems. 
We will address these issues in more detail with future papers on high-resolution 
metal-line observations complete with \nhi\ measurements.

Also note that most systems presented here are not confirmed 
DLAs as they lack spectral coverage of \lya, but
(as will be shown) 
the following analysis 
is largely independent of \ion{H}{1} measurements.  
Previous works (eg.\ Khare et al. 2004)
have attempted to estimate $\mnhi$ from the reddening $E(B-V)_{g-i}$.
But, we find that for the systems in our sample where \nhi\ values
exist, this method 
consistently overestimates $\mnhi$, often dramatically so.  This may be 
due to reddening contributions from both the QSO host and the absorber
in question, 
as well as possible differences between the properties 
of local and high-redshift dust 
grains.  We therefore avoid these rough
estimations for $\mnhi$ and await \lya\ observations to comment on ionization corrections and 
dust-to-gas ratios of particular systems. 
A quantitative justification of the term `MSDLA' is provided below in $\S$~3.2.

\section{Metal Strong Absorbers in the SDSS}

The SDSS is a tremendous survey conducted using a 
2.5-meter telescope at the Apache Point Observatory (APO, Sunspot, NM).  Millions 
of objects have been observed by this wide-field 
digital telescope. All of the SDSS spectra analyzed here were
reduced using the SDSS spectrophotometric pipeline.  
The third dataset, SDSS-DR3, 
contains all data taken through June 2003, and we retrieved the quasar 
spectra from http://www.sdss.org.  
With rare exception, the fiber-fed SDSS spectrograph 
provides full-width half-maximum FWHM $\approx 150$ km/s spectra 
of each quasar for the wavelength range 
$\lambda \approx 3800 - 9200$\AA.  The $1\sigma$ 
Poisson noise from counting statistics 
is also calculated and recorded during the reduction.  
For our metal-strong subset, we chose a limiting 
magnitude of $r=19.5$ mag to facilitate follow-up observations 
with 10m-class telescopes.
The signal-to-noise ratios (SNRs) of this subset of SDSS spectra range from 
$\approx$ 5 to 30 per pixel, with a typical value of $\approx$ 12.  

\begin{figure}[ht]
\centerline{
\includegraphics[angle=90,width=3.4in]{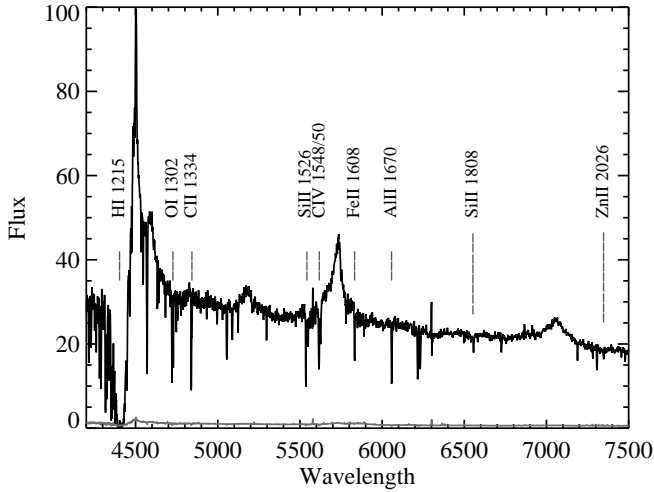}}
\caption{Sloan Digital Sky Survey spectrum of DLA-B/FJ0812+32 
(plate 861, fiber 333), 
with wavelengths in \AA\ along 
the x-axis, flux ($f_{\lambda}*10^{17} {\rm erg \, s^{-1} \, 
\cm{-2} \, \AA^{-1}}$) along the y-axis, and the 
$1\sigma$ error in gray.  This example shows a bit of 
the \lya\ forest between 4200-4450 \AA, the damped \lya\ profile
of log $\mnhi=21.35$ near 4400 \AA\ 
($z_{abs}=2.626$), and QSO \lya\, NV, \ion{Si}{4} \& OIV, and \ion{C}{4} 
emissions near 4500, 4600, 5200, 5700 and \AA, respectively ($z_{em}=2.701$).  
Various strong metal absorption lines
can be seen scattered redward of the QSO \lya\ emission peak, many of 
them associated 
with the DLA system 
(marked; the longer dashes show the key metal-strong transitions 
\ion{Si}{2} 1808 and \ion{Zn}{2} 2026).  
This is a bright QSO ($r=17.46$ mag) and therefore this 
spectrum has high SNR for SDSS data.
\label{861333}}
\end{figure}

Figure~\ref{861333} shows 
the SDSS spectrum of the metal-strong DLA system first 
identified by \cite{nature03}.  
The successful detection of more than 20 elements in this DLA 
(DLA-B/FJ0812+32) motivated us to develop an automated procedure to 
identify a complete catalog of similarly metal-strong systems in SDSS.

\subsection{SDSS Search Algorithm and Visual Inspection}

This section describes the 
codes we developed to identify metal-strong DLA candidates in the SDSS quasar
spectra. These
algorithms are now available as part of the 
XIDL package developed by J.X. Prochaska; see 
http://www.
\\ucolick.org/$\sim$xavier/IDL/index.html.  
Our strategy was to identify all significant absorption features
redward of the \lya\ forest and search for sets of these lines which have
a common absorption redshift.
The search is complicated by the (possible) presence of 
other metal lines associated with separate systems 
along the same sightline as well as telluric lines from our atmosphere.  

We set the minimum redshift of this survey to $z_{abs}=1.6$ because  
the Earth's ozone (${\rm{O}}_3$) layer blocks UV 
radiation below 3000~\AA\ 
and, therefore, the \lya\ profile cannot be observed at $z_{abs}<1.6$ with 
ground-based telescopes.  
Follow-up UV observations are largely unattainable because of the current lack 
of space-based UV facilities with the requisite wavelength
coverage and resolution.
SDSS spectra have a starting wavelength at $\approx 3800$\AA\ such that
systems with $z_{abs}<2.1$ 
have their \lya\ profile blueward of the SDSS spectra.  
It is nevertheless worthwhile to extend the search to 
$z=1.6$ rather than $z=2.1$, because 
we expect to observe more metal-strong systems at 
lower redshift given: (1) galaxies enrich in time 
\citep{pro03a}; and (2) the QSO luminosity 
function peaks at $z_{em} \sim 2$ providing many additional targets at 
$z < 2.2$.  Also, extending the search to $z=1.6$ is 
worthwhile simply because we'd like to increase our chances of 
of finding potentially rare metal-strong DLAs.
Many of our metal-strong candidates are detected at 
$z_{abs}<2$ and therefore have no corresponding \lya\ profile in the SDSS spectra.  
Follow-up observations with medium-sized ground-based telescopes will provide the 
\ion{H}{1} measurements necessary for determining metallicities and ionization fractions of the systems.

The first step in our analysis is to process every quasar with the
algorithm {\it{sdss\_qsolin}}. 
This routine fits a continuum to each spectrum using
a principle component analysis (PCA) developed by S. Burles.
It then convolves each spectrum with a Gaussian of FWHM$=$2.5 pixels 
(chosen to match the width of unresolved metal profiles in SDSS)
and records any 
resulting absorption features that are detected to be $\ge 3.5\sigma$.  
Finally, the code records the wavelengths of all features separated 
by more than 4 pixels; multiple features identified with smaller separation are 
recorded as a single line. 
We restrict the metal search 
to the spectral region redward of the QSO \lya\ emission 
to (1) avoid misidentifying \lya\ forest lines as metal absorption, and (2) 
simplify the automated continuum fitting.  We estimate that a negligible amount 
of candidate metal-strong systems are missed by avoiding the \lya\ forest in our
metal search, largely due to the possibility of 
lone \ion{Mg}{2} systems being retained by the algorithm, as well as 
the broad $\lambda_{rest}$ range of other typically strong lines used in the 
search (see Table~\ref{tab:fosc}).  Also note that this issue is particularly 
negligible
when considering the MSDLA fraction computed in $\S$~7.1, due to the availability 
of \lya\ absorption and high $\lambda_{rest}$ transitions (eg.\ \ion{Mg}{2} \& 
\ion{Fe}{2}).

The wavelengths of the detected features are then 
sent to {\it{sdss\_search}} (the controlling code for 
{\it{sdss\_metals}}, 
\\{\it{sdss\_compare}} and {\it{sdss\_dla}}, the latter 
detailed in Prochaska \& Herbert-Fort 2004) where they are matched against 
14 redshifted metal transitions (see Table~\ref{tab:fosc}).  
These 14 transitions were chosen due to their frequent occurrence in 
high-redshift DLAs; the ions are typically present in large amounts 
(due to SNe enrichment 
from massive stars) and/or have 
large oscillator strengths.  We systematically 
redshift the 14 lines from $z=1.6$ to $z_{em}$ 
in increments of 0.0001 in $z$.     
For each corresponding absorption feature in the observed spectrum 
within the local dispersion $\dell$ of the spectrograph 
(roughly 1 \AA\ pixel$^{-1}$ 
near 4000 \AA\ to 2 \AA\ pixel$^{-1}$ near 8000 \AA), 
the algorithm records a match.
The matched wavelengths are also restricted to be $>1230 \times $(1+$z_{em}$) to 
avoid misidentifying QSO-associated \lya\ absorption as metal absorption, and 
regions of the spectrum containing severe atmospheric effects (particularly 
in the red) are avoided.  
If more than 10 
lines were identified redward of 8000 \AA, this region is flagged for 
severe sky effects and is not included in any further analysis.
Otherwise, the search terminates at 9200 \AA.  

At each redshift, we calculate a percentage-based detection to be the ratio of matches to the 
number of possible matches; $P=n/m$, where $n$ is the number of 
matched transitions and $m$ 
is the number of possible matched transitions, specifically those lying in the spectral
coverage (excluding the \lya\ forest) and not in a masked sky area.  
We are especially interested in detecting the \ion{Si}{2} 1808 transition (after 
examining many SDSS spectra we have determined that it is an excellent metal-strong 
indicator; see $\S$~5 below); therefore, we do not include it as a 
possible match (the $m$ of our percentage-based detection), but do increment
$n$ if this line is detected.  
We then define candidate absorption 
systems to be at those redshifts where $P>60\%$ and $n>1$.  
If two or more of these systems lie within $\delz = 0.01$ we 
combine them into one system (to account for wide absorption line systems 
and the low resolution of SDSS spectra).
Finally, the redshift, the percentage of detection and number 
of line matches, $n$, of each detected metal system are recorded.  
The latter two are then used to assign each system an overall quality rating in {\it{sdss\_compare}}. 


The final step was to visually inspect every detected system 
with a customized tool, {\it{sdss\_finchk}}, 
and visually rate the strengths of the metal absorptions.  
Each system was subjectively rated as either `bizarre', `none', `weak', 
`medium', `strong', or `very strong', 
depending on the 
\\amount and strengths of the lines 
present in the system.  
We mostly used the presence 
of \ion{Si}{2} 1808 to judge `strong' and `very strong' systems in SDSS, as \ion{Zn}{2} 
may often be buried in the noise of low-SNR SDSS spectra.  
If the minimum
depth of the \ion{Si}{2}\,1808 profile
$F_{min}/F_{q} \approx 0.9$ where $F_q$ is the quasar flux, then the 
system was rated as `strong'.  
If the normalized intensity was $F_{min}/F_q \le 0.85$ the system was 
rated `very strong'.  
Note that this is a subjective visual inspection 
and the minimum depths were chosen based on experience of looking at many 
SDSS \ion{Si}{2}~1808 absorption profiles.
Any other metals present were also taken into 
account, especially if the \ion{Zn}{2} 2026 line was covered.  
Candidates that were found as a result of confusion with 
severe noise or sky lines were rated `bizarre'.  For reference, a summary of our subjective 
metal-rating scale is shown in Table~\ref{mtl_rat}.

All 435 `strong' and `very strong' candidate systems 
from our search in SDSS-DR3 ($z_{abs} \ge 1.6$) are compiled 
in Table~\ref{all_DR3}.  
The complete table is available in the electronic (online) version 
of this paper; we 
present only a sample here. 
Table~\ref{all_DR3} lists SDSS plate, MJD, and fiber, together with 
RA, Dec, $r$, $z_{em}$, $z_{abs}$, the 
overall quality rating of each system (18 is the highest with a strong candidate 
DLA automatically detected by {\it{sdss\_dla}}, otherwise 10 if 
the system lacks \lya\ coverage)
and our metal rating from visual inspection (4=`strong', 5=`very strong'). 
Most ($\frac{17}{27} \approx 63 \%$) systems in our ESI sample 
(described below) were taken from the `very strong' category, while 
the remainder are all classified as `strong'. 
Broad absorption line (BAL) spectra \citep[see][]{barlow94} 
were avoided if determined to 
be too severe (via subjective visual inspection), 
as these are often associated with the QSO itself and can 
significantly confuse any subsequent absorption analysis.  Approximately
3\% of the SDSS-DR3 quasar sample were flagged as BALs (PHW05). 

\subsection{SDSS Search Results}

Of the 19,435 SDSS-DR3 QSO sightlines with $z_{QSO} \ge 1.6$, 
2,352 systems show metal absorption 
ranging from `weak' to `very strong' (ie.\ all but the 
'bizarre' and 'none' categories).  
16,649 sightlines (86\% of those searched) were 
without sufficient features resembling DLA and/or metal absorption to 
be retained by the algorithm (note that, for example, a 
single \ion{C}{4} absorption system won't satisfy our search criteria).  
From the set of 
2,352 candidate metal absorbers (not yet limited to $r < 19.5$), 
we rated 285 (12\%) 
as `strong' (S) and 150 (6\%) 
as `very strong' (VS) systems (see Table~\ref{all_DR3}).  
Of the 78 systems categorized as `strong'
with $z_{abs} \ge 2.2$, 74 (95\%) show a corresponding DLA, while 
3 of the 4 others have \nhi\ within $1\sigma$ of the DLA 
threshold.   The exception at $z=2.44$ toward 
J150606.82+041513.1 (SDSS plate and fiber 
[589,547]) has a measured log$\N{HI}=19.65 \pm 0.15$ at $z_{abs}=2.44$ and 
represents a rare subset of the super-LLS population. 
Therefore, we report that $< 5 \%$ 
of S absorbers in our sample 
(with observable \lya\ profiles in SDSS) are not DLAs.    
Of the 41 VS systems 
with $z_{abs} \ge 2.2$, 100\% show a corresponding DLA in the SDSS spectra.  
Even allowing for an evolution in the \nhi\ distribution of 
metal-strong candidates between redshifts 
$1.6 < z < 2.2$, we contend that 
only a very small fraction of our MSDLA candidates are not 
truly DLAs.  We are therefore confident in 
having identified a metal-strong DLA population 
while lacking \lya\ coverage on most systems.  
However, the presence of a larger fraction of lower \nhi\ systems cannot be 
excluded until \nhi\ values are obtained for all systems in our sample.

Note, however, that we do have one confirmed super-LLS case in our 
sample -- discussed 
below; also see \cite{peroux06} for another possible example of such a system, 
although note that the \nhi\ value of the \cite{peroux06} system 
was within $0.5\sigma$ of the DLA threshold, and an independent analysis of 
the same data from \cite{Rao05} found it to be a 
bona fide DLA with log$\N{HI} = 20.54 \pm 0.15$.  Other systems with low \nhi\ and 
high metallicity have also been found at lower redshifts ($z < 1$) 
by \cite{pettini00}
and Jenkins et al.\ (2005; note that this 
system lies at $z \approx 0.08$, ie.\ far below the range studied in this work).

\clearpage

\section{ESI Observations, Data Reduction and Measurements}

\subsection{ESI Observations and Data Reduction}

To test our automated selection method we compiled a list of our 
strongest candidate metal absorption systems from SDSS and acquired \ndla\ 
moderate-resolution spectra using ESI on UT December 20th, 2003, 
and September 10th and 11th, 2004, at the 10m Keck II telescope.  
The SNRs of our ESI 
spectra range from $\approx 10-20$ per pixel with a typical value 
of $\approx 16$.  ESI has a 
pixel size of $\approx 11$ \kms, and a $0.5''$ 
slit covers 3 pixels for a FWHM of $\approx 34$ \kms.  
The wavelength coverage is roughly 4,000 \AA\ - 
10,200 \AA.  Table~\ref{tab:obs} 
presents a log of our observations listing 
QSO name, SDSS plate, MJD and fiber, QSO emission redshift ($z_{em}$), 
$r$ magnitude, exposure time, slit width, and observation date.  

The data were reduced with the ESIRedux software package 
(Prochaska et al.\ 2003a; see http://www2.keck.hawaii
\\.edu/inst/esi/ESIRedux/).
This package converts 2D 
\\echelle spectra into 1D, 
wavelength-calibrated spectra.  
The $1\sigma$ array is also calculated during the reduction process.
We continuum-fit each QSO separately with custom software, {\it{x\_continuum}}, 
by fitting 
high-order polynomials to separate pieces of the spectrum 
containing no significant absorption.  The fit pieces are patched 
together to create a smooth trace of the QSO continuum.
We caution the reader that no errors from our continuum-fitting 
are taken into account. This is a significant source of error when measuring 
very weak lines and so we report many such cases as upper limits.  
Note that continuum error will be comparable to the statistical error 
for $<4\sigma$ detections but negligible otherwise.

\subsection{Measurements from ESI data}

\begin{figure}[ht]
\centerline{
\includegraphics[width=3.5in]{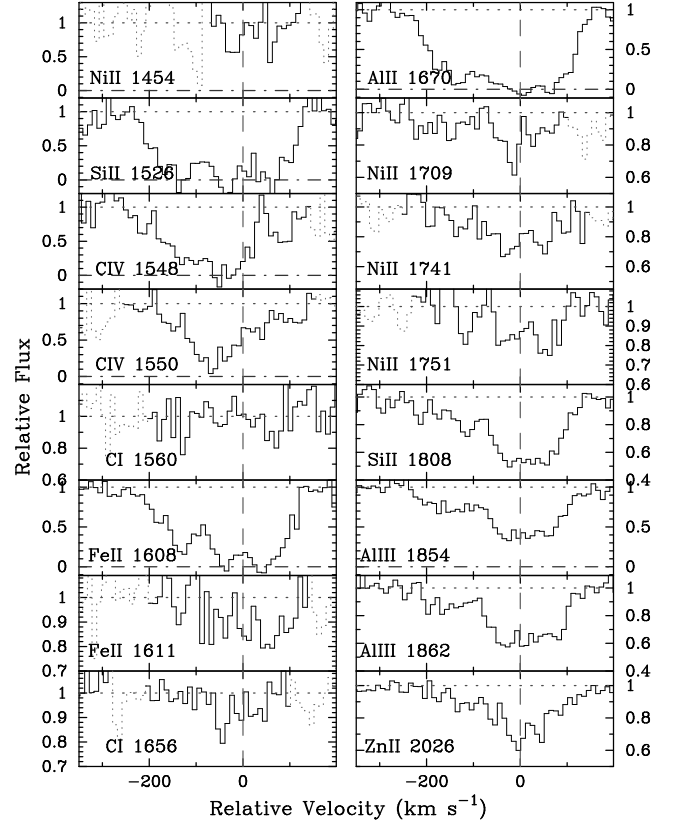}}
\caption{Sample ESI velocity plot 1, 
with $v=0$ corresponding to the redshift of the 
absorption system (SDSS0008--0958, $z_{abs} = 1.768$).  The dashed 
lines trace the normalized continuum.  
Line blending is indicated with dotted lines and were 
not included in the measurements.  
Note the structure seen in unsaturated lines.  
\label{vel1}}
\end{figure}

\begin{figure}[ht]
\centerline{
\includegraphics[width=3.5in]{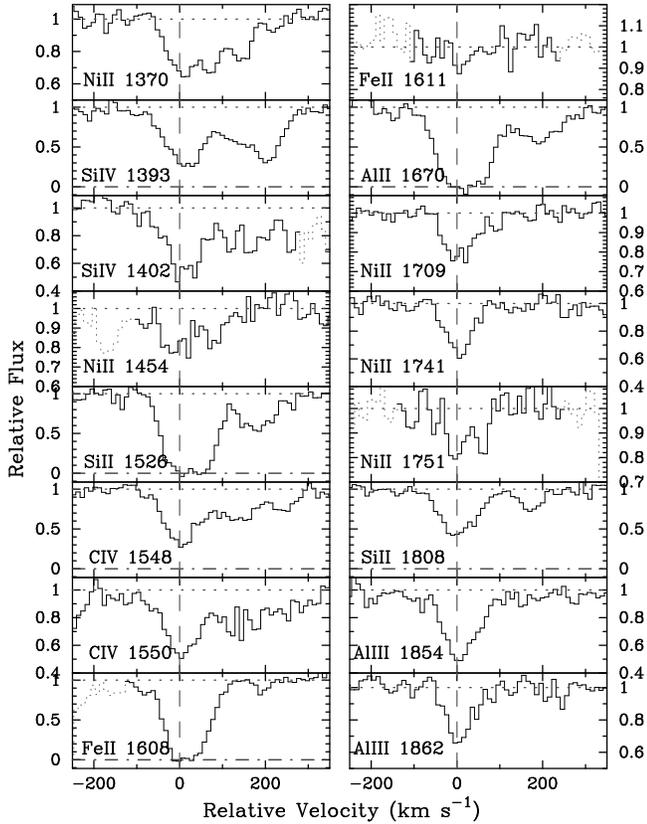}}
\caption{Sample ESI velocity plot, same as Figure~\ref{vel1} but
for SDSS0844+5153 at $z=2.77489$.
\label{vel2}}
\end{figure}

Ionic column densities are 
determined using the apparent optical depth method (AODM; 
Savage \& Sembach 1991, also Jenkins 1996), except when determining 
\\$N({\rm{ZnII\ 2026}})$ and $N({\rm{ZnII\ 2062}})$ (discussed below in 
\\$\S$~4.2.2).  
Only lines that have been detected at $\ge 3\sigma$ are listed as measurements.
We begin a measurement by plotting the continuum-normalized profile 
in velocity space with an 
arbitrary zero-velocity centered on the redshift $z_{abs}$ of the system 
(usually determined from the strongest of 
all visible transitions).  As examples, 
selected velocity plots from two systems are shown 
in Figures~\ref{vel1} and \ref{vel2}.
Our full ESI velocity plot sample can be retrieved from 
the electronic edition of the journal.
If an absorption line is determined to 
be saturated (equivalent width $W>600$m\AA,; see \citealt{pro03b} and below), 
$\N{X}$ may be treated as a lower limit to 
the true column density.  However, note that saturation effects based 
on $W$ measurements alone may be misleading for systems with wide velocity 
widths, as is clearly the case with SDSS0016--0012 (integrated 
velocity width $\approx 1000$ \kms wide).  See below for more discussion of saturation 
effects and its impact on MSDLA classification for Keck ESI data. 
The $\N{X}$ values 
are reported as  upper limits if a feature is detected at 
less than  $3\sigma$ statistical significance.

The electronic edition of the journal also 
presents tables of ionic column densities for each QAL system in our 
metal-strong sample. 
Each table lists ions, rest wavelengths in \AA, a flag (0-5) distinguishing primary/non-primary ions 
and/or limits, $N(\rm{ion})$ 
as determined from the AODM and
$N(\rm{element})_{adopt}$ (determined to be the weighted mean of
the primary ion column densities).
Any solar values used in the analysis are taken from \cite{grvss96}. 

\subsubsection{Saturation of ESI spectra and the MSDLA definition}

Here we emphasize the importance of saturation effects 
in Keck ESI spectra.  We find (in this study 
and from previous experience with ESI data) that absorption profiles extending below 
normalized intensities $F_{min}/F_q \le 0.4$ are typically saturated and must be treated as lower limits to the true column densities.  In light of this,  we 
treat systems with saturated \ion{Zn}{2} or \ion{Si}{2} profiles above 
log$\N{Zn^+} \ge 13.0$ or log$\N{Si^+} \ge 15.8$ as MSDLAs; however, 
note that true MSDLAs are defined to have log$\N{Zn^+} \ge 13.15$ 
or log$\N{Si^+} \ge 15.95$ and we are simply accounting for ESI saturation 
effects.
Indeed, DLA-B/FJ0812+32 
clearly illustrates this effect; ESI data show log$\N{Zn^+} \approx 13.04$ while KeckI/HIRES 
yields log$\N{Zn^+} \approx 13.15$ because the lines are not fully
resolved in the ESI spectrum.
Also, ESI data of $\N{Si^+}$ from DLA-B/FJ0812+32 shows log$\N{Si^+} 
\\> 15.78$ whereas 
HIRES data yields log$\N{Si^+} \approx 15.95$, again due to line saturation.
Also note the possibility that some very narrow saturated components 
of these transitions may not even be 
resolved by HIRES, in which case the true column densities could be even higher, 
and so some systems below the ESI-saturation MSDLA threshold might also 
truly be MSDLAs.  
Such cases could raise the MSDLA fraction higher than is reported below ($\S$~7.1).

\subsubsection{\ion{Mg}{1} and \ion{Cr}{2} contributions to \ion{Zn}{2}}

\cite{pettini94} emphasized (for the DLAs) that the \ion{Zn}{2}~2026
profile is blended with a weak yet potentially important
\ion{Mg}{1}~2026 transition.  \cite{york06} 
have recently shown that in some cases 
(eg.\ their Sample 21) that \ion{Mg}{1} may 
dominate the absorption profile at 
$\lambda_{rest} = 2026$\AA\ (as we also find in a few systems), 
causing concern for any study examining \ion{Zn}{2} abundances, 
particularly for strong systems.   
Although the \ion{Mg}{1}~2026 line has not been well surveyed in the damped \lya\
systems \citep{pw02}, we suspect that this contribution (in MSDLA 
candidates) is significantly larger
than the average.  
The likely explanation is that the metal-strong absorbers
correspond to higher density regions (sightlines probing nearer 
the enriched central regions of galaxies) and so one observes a higher fraction of
Mg$^0$ atoms per Zn$^+$ ion.

Because our observations generally include the \ion{Mg}{1}~2852 transition,
we can estimate the equivalent width of the \ion{Mg}{1}~2026 transition 
(assuming the linear curve of growth, or COG) from the column 
density measured for \ion{Mg}{1}~2852 (with the AODM) 
and therefore its contribution to the line-profile 
at $\lambda_{rest} = 2026$ \AA.  In turn, we can 
measure a more reliable value for $\N{Zn^+}$ 
from the equivalent width of the remaining \ion{Zn}{2} transition at 
$\lambda_{rest} = 2026$ \AA\ (again assuming the linear COG).
An inspection of Table~\ref{tab:znanly} shows that the equivalent width
of the \ion{Mg}{1}~2026 line often contributes $> 20\%$ 
(with a large scatter) of the
total equivalent width measured at $\lambda_{rest} = 2026$ \AA.  For 
those cases where we expect a saturation correction for \ion{Mg}{1}~2852, 
we have incremented the column density by 0.1 dex.  This rather
small adjustment is due to the fact that the line-profiles are generally
not heavily saturated (specifically, follow-up observations with HIRES
indicate a 0.1 to 0.2\,dex correction is appropriate given the observed
peak optical depth of these lines at ESI resolution).
We caution 
that this correction may be misleading for the strongest 
\ion{Mg}{1} systems and that 
such cases could lead to an overestimate of $\N{Zn^+}$.  Higher-resolution 
data will be 
useful for examining this effect in particular systems.  
Furthermore, we have visually 
inspected every \ion{Zn}{2}~2026 profile and 
find no evidence for severe \ion{Mg}{1}~2026 contamination
for those DLA where we report a $\N{Zn^+}$ value.  Also
note that in many cases 
(see Table~\ref{tab:znanly}) the profile of this transition 
is either noisy (due to low SNR in a particular section of our echelle data), 
blended 
with sky and/or profiles from unrelated systems, or simply weak and perhaps 
at $< 3\sigma$ statistical significance if considering continuum-placement errors, and we report $\N{Zn^+}$ as an upper limit.  

Similar to the \ion{Mg}{1} blending issue is the blending of 
\ion{Cr}{2} and \ion{Zn}{2} at $\lambda_{rest} = 2062$ \AA.  This effect was 
estimated in the 
same manner as with \ion{Mg}{1} (above; yet with both \ion{Cr}{2} transitions at  
$\lambda_{rest} = 2056$ and 2066 \AA\ whenever possible) 
and we calculate an independent 
value of $\N{Zn^+}$ at $\lambda_{rest} = 2062$ \AA, whenever possible.  
We find that 
\ion{Cr}{2} significantly dominates the 2062 profile in most cases. 
When taken together we find (and list in the tables and use in all plots) 
the \ion{Mg}{1} and \ion{Cr}{2} {\it{blend-corrected}} values of $\N{Zn^+}$.  
Nevertheless, one must always 
keep in mind the limited resolution of the ESI and that equivalent width
measurements may also underestimate $\N{Zn^+}$.

\section{SDSS Metal-Strong Indicators}

Having identified all potential metal-strong systems in SDSS-DR3, we determine rest
equivalent widths $W$ for transitions detected in both the SDSS and ESI spectra.
$W$ is the normalized, intensity-weighted width of a line (here in m\AA), and 
corresponds to the fractional energy absorbed by the transition.  This quantity is independent 
of instrument resolution and is a good predictor of 
column density for weak lines.  Strong lines tend to be saturated; in this regime, $W$ 
grows with log$\N{X}$ and so a line has roughly the same $W$ as $\N{X}$ 
increases.  As a result, $W$ is not a reliable column density 
predictor for strong lines and we must identify the most reliable weak lines 
as metal-strong indicators in low-resolution spectra.

\begin{figure}[ht]
\centerline{
\includegraphics[angle=90,width=3.4in]{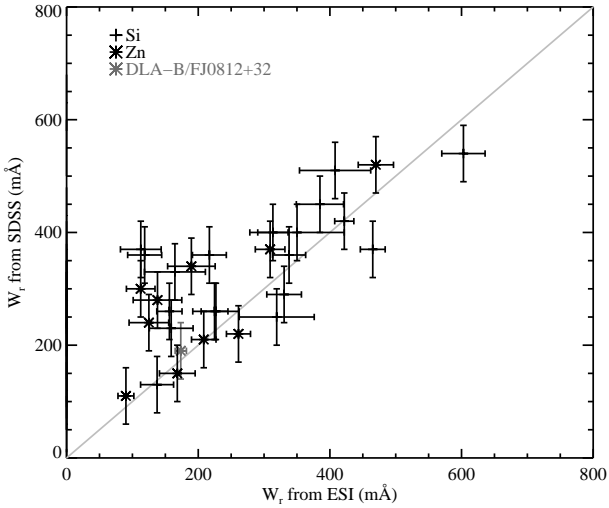}}
\caption{Rest equivalent width values $W_r$ 
from ESI vs.\ $W_r$ from SDSS for both \ion{Si}{2} 1808 and \ion{Zn}{2} 2026, 
for all cases with secure column density values.  
The value for DLA-B/FJ0812+32 is plotted in gray.
Note that these values do not account for minor blends, low SNR or saturation effects; weaker 
$W_r$ lines systematically have higher, false contribution from noise in lower-resolution and 
lower-SNR (SDSS) data.
\label{WW_both}}
\end{figure}

We have determined that \ion{Si}{2} 1808 and \ion{Zn}{2} 2026 are usually 
evident in the metal-strongest absorption systems in SDSS.  
These two elements are typically only mildly depleted, 
making them useful indicators of MSDLAs.  
We measured $W_r$ values 
for \ion{Si}{2} 1808 and \ion{Zn}{2} 2026 in both the SDSS and ESI 
spectra for the subset of systems observed with ESI.  
However, recall that the \ion{Zn}{2} 2026 transition is typically blended with 
\ion{Mg}{1} 2026. 
Table~\ref{tab:ind} lists the QSO name, 
SDSS plate and fiber, $z_{em}$, $z_{abs}$, 
$r$, log$\mnhi$ if available, log$\N{SiII\ 1808}$, 
log$\N{ZnII\ 2026}$ (blend-corrected), 
and $W_r(1808)$ and $W_r(2026)$ from both SDSS and ESI data.  
The idea here is to roughly estimated $W_r$ values in SDSS data to gauge a
system's metal-strong potential when observed at higher resolution.
We assign conservative $\sigma_W$ estimates on SDSS data  
of 50 m\AA. 
Figure~\ref{WW_both}
shows a plot of our $W_r$ measurements for the absorption 
profiles at $\lambda = 1808$ and 2026\AA\ from SDSS and
ESI observations, for systems with secure column density values.  
The majority of measurements lie within $2\sigma$ of each other, 
although weaker lines are found to have 
a systematically higher, false contribution 
from noise in lower-resolution and lower-SNR SDSS data.  Overall, however, 
we conclude that using $W_r(1808)$ and $W_r(2026)$ from SDSS 
is a reliable means of gauging a system's metal-strong 
potential in higher resolution spectra.

\begin{figure}[ht]
\centerline{
\includegraphics[angle=90,width=3.4in]{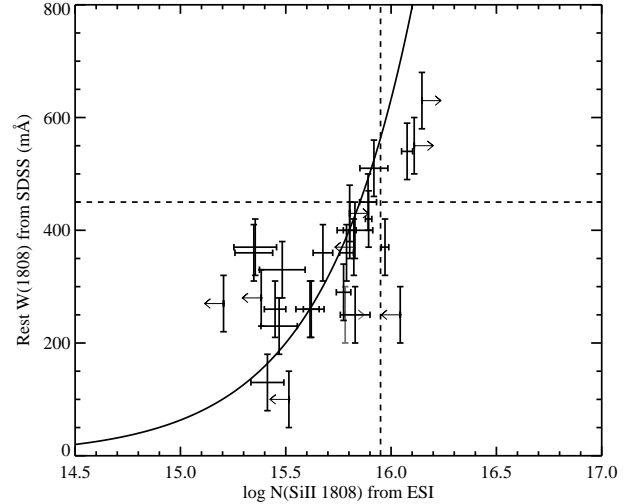}}
\caption{
$W_r(1808)$ from SDSS vs.\ log $\N{Si II\ 1808}$ from ESI.  
DLA-B/FJ0812+32 is 
plotted in gray as a lower limit to $\N{Si^+}$.   
Also plotted in black dashed linestyle 
are the proposed log$\N{Si^+}$ and $W_r(1808)$ metal-strong thresholds.
Note that ESI values may be underestimated for log $\N{Si II 1808} > 15.8$ 
due to possible saturation.  
The solid black 
line traces $W_r(\rm{Si II\ 1808})$ vs.\ $\N{Si II\ 1808}$ for optically-thin gas.
\label{WNSi}}
\end{figure}

\begin{figure}[ht]
\centerline{
\includegraphics[angle=90,width=3.4in]{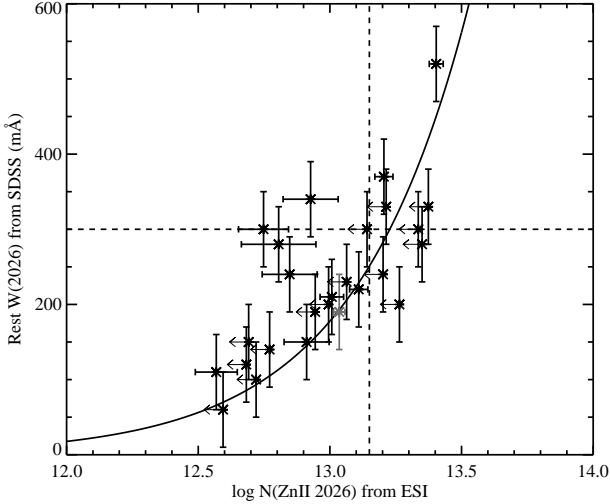}}
\caption{
$W_r(2026)$ from SDSS vs.\ log $\N{Zn II\ 2026}$ from ESI.  
DLA-B/FJ0812+32 is plotted in gray.   
Also plotted in black dashed linestyle 
are the proposed log$\N{Zn^+}$ and $W_r(2026)$ metal-strong thresholds.
Note that ESI values may be underestimated for log $\N{Zn II 2026} > 13.0$ 
due to possible saturation.  
The solid black 
line traces $W_r(\rm{Zn II\ 2026})$ vs.\ $\N{Zn II\ 2026}$ for optically-thin gas.
\label{WNZn}}
\end{figure}

We now investigate how $W_r$ estimates from SDSS correlate 
with column density values measured from the ESI data.  
Figures~\ref{WNSi} and \ref{WNZn} 
show $W_r$ from SDSS vs.\ log$N$ from ESI for  \ion{Si}{2} 1808 and  
\ion{Zn}{2} 2026, respectively 
(exact $\sigma_N$ values can be found in the abundance tables of the electronic 
version of the paper; 
roughly 0.07 dex for  \ion{Si}{2} 1808 and 0.10 dex for  \ion{Zn}{2} 2026).  
We expect smaller $W_r$ 
values to be more reliable indicators of $N$ (they are weak, unsaturated lines), 
and indeed we see a larger scatter 
in the larger and possibly saturated $W_r(\rm{Si\ II\ 1808})$ values. 
We therefore expect that  \ion{Zn}{2} 2026, although less often detected in 
SDSS spectra, is a more reliable indicator 
of potential MSDLAs than is \ion{Si}{2} 1808. 
Considering Figure~\ref{WNZn}, we tentatively expect that most 
SDSS systems with $W_r(\rm{Zn\ II\ 2026}) > 300$ m\AA\ will be truly 
metal-strong when measured at higher resolution, supporting a choice to 
skip the moderate-resolution 
confirmation observations of such systems.  Note that our ESI 
measurements of \ion{Zn}{2} lines near log$\N{Zn\ II\ 2026} \approx 13.0$ are 
systematically underestimated by 0.1 to 0.2 dex due to line saturation.
 \ion{Si}{2} 1808, however, is detected more often than 
 \ion{Zn}{2} 2026 in the SDSS spectra 
and may be the only indicator for low-resolution, low-SNR data.
We therefore propose a similar tentative 
$W_r(\rm{Si\ II\ 1808})$ metal-strong threshold 
of 450 m\AA, 
past which we will skip medium-resolution observations (also note here 
that our ESI measurements of \ion{Si}{2} lines near log$\N{Si\ II\ 1808} 
\approx 15.8$ may be 
viewed as underestimates due to line saturation).
For optically-thin gas, a $W_r(\rm{Zn\ II\ 2026}) = 300$ m\AA\ profile corresponds to 
log$\N{Zn^+} \approx 13.2$, and a $W_r(\rm{Si\ II\ 1808}) = 450$ 
m\AA\ profile corresponds to a log$\N{Si\ II\ 1808} \approx 15.9$.  
With the overall goal of this project being to discover more systems like DLA-B/FJ0812+32 
we are excited to see that many systems far surpass it in $\N{Si^+}$ and $\N{Zn^+}$.  
These systems may show transitions not yet observed in the young universe and 
could be used to constrain theories of nucleosynthesis and galaxy evolution, as did 
DLA-B/FJ0812+32.  We will present detailed results of KeckI/HIRES and 
VLT/
\\UVES observations of these new systems in upcoming papers.

\section{Equivalent Width and SNR Estimates for Detecting Very Weak Lines}

To reliably constrain theories on the 
production of elements like B, O, Sn, Pb and Kr, 
higher-resolution data is needed.  
Some of these lines (BII $\lambda$1362, \ion{O}{1} $\lambda$1355, 
SnII $\lambda$1400, \ion{Pb}{2} $\lambda$1433, and 
KrI $\lambda$1235) are still out of reach to modern 
instruments and with reasonable integration times.  
Nevertheless, it is interesting to estimate how strong they might be in 
our sample of metal-strong 
DLA galaxies for which we have reliable measurements of 
other elements.  These extremely weak 
lines are presumed to lie buried 
in our metal-strong ESI data, yet it may 
be possible to bring them out with lengthy HIRES observations (assuming that they lie 
in detectable spectral regions).  As a brief exercise we predict the SNRs it would take to 
detect such features at $3\sigma$.

Because we are interested here 
in detecting weak lines, we will make use of the weak limit of 
equivalent width,

\begin{equation}
W_{obs}({\rm X}) = {\pi e^2 \over m_e c^2} \lambda_r^2 (1+z) N({\rm X}) f \;\;,
\label{eqn:W}
\end{equation}

\noindent where $W_{obs}(\rm X)$ is the {\it{observed}} equivalent 
width of transition X and
$\N{X}$, $\lambda_r$ and $f$ are the column density,
rest wavelength, and oscillator strength of the 
line, respectively.  
Because we have not yet measured the weak lines, we must first 
estimate their $N$ values from other reliable line measurements in their corresponding
systems.  To do this, we assume that the column densities scale with 
their solar abundances (ie.\ [X/Y]=0 with X the 
undetected element and Y a detected reference element) 
{\it{with no corrections}}.  
Next, a reliable reference element is chosen.  Because Fe is highly refractive, we will 
choose the mildly depleted element Si (we have 
the most measurements of Si after Fe).  Using 

\begin{equation}
\log{[N({\rm X})/N({\rm{X}})_{\sun}]} = \log{[N({\rm Si})/N({\rm{Si}})_{\sun}]}
\label{eqn:estB}
\end{equation}

\noindent to scale the abundances to solar values, one can easily solve for $\N{X}$ 
and hence $W_{obs}(\rm X)$.

To determine what SNR is needed for a $3\sigma$ detection, it can be shown that

\begin{equation}
{\rm{SNR}}=\frac{3}{W_{obs}}\sqrt{{\frac{{\rm V}\lambda_{obs}\dell}{c}}} \;\;,
\label{eqn:snr}
\end{equation}

\noindent assuming $\sigma_i$ (the normalized error of the spectrum) 
and $\dell$ are constant across the V velocity width feature.  This is a reasonable assumption 
for metal lines in the weak limit (and if they are narrow features). 
Here $\dell$ is the HIRES dispersion element in m\AA/pixel at the wavelength of the 
feature 
and $c$ is the speed of light.

\begin{figure}[ht]
\centerline{
\includegraphics[angle=90,width=3.4in]{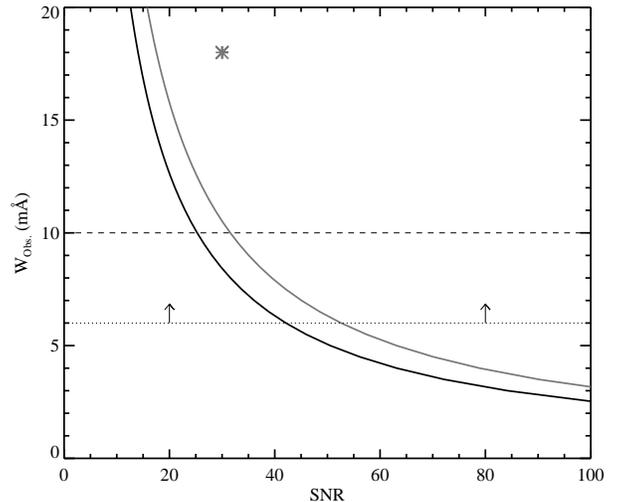}}
\caption{Limiting $3\sigma$-detection lines of $W_{obs}$ vs.\ SNR 
(per 2 \kms\ pixel) 
for a V = 20 \kms\ profile observed with HIRES at 4000\AA\ (black) 
and 5000\AA\ (gray).  
The dotted line shows $W_{obs}$ of \ion{B}{2}\,1362 
in DLA-B/FJ0812+32 as predicted from 
ESI data ($W_{obs} > 6$ m\AA; a lower limit because the reference 
element used, Si, is a lower limit in the ESI data, log$\N{Si^+} > 15.8$).  
The dashed line is the predicted $W_{obs}$ of \ion{B}{2}\,1362 
using the log$\N{Si^+} = 16.0$ measurement from HIRES as reference.  
In the HIRES spectrum, we measure the $W_{obs} \approx 18$ m\AA\ 
for a SNR $\approx 30$ (per 2 \kms pixel), marked here as the gray star.
\label{WSNR}} 
\end{figure}

Figure~\ref{WSNR} shows the limiting $W_{obs}$ detection 
lines vs.\ SNR (per 2 \kms pixel) for a V = 20 \kms absorption profile 
detected at $3\sigma$ and plotted for $\lambda_{obs}=4000$ \AA\ (black) 
and 5000 \AA\ (gray).  
As an example, we investigate predictions for the observed \ion{B}{2} 1362 
line in the DLA-B/FJ0812+32 HIRES spectrum from 
\cite{nature03} (V $\approx 20$ \kms, $\lambda_{obs}=4940$ \AA).  
The dotted line shows $W_{obs}$ of \ion{B}{2} in DLA-B/FJ0812+32 as predicted from 
ESI data assuming [B/Si] = 0 ($W_{obs} > 6$ m\AA; a lower limit because 
log$\N{Si^+} > 15.78$ in the ESI data).  
The dashed line is the predicted $W_{obs}$ of \ion{B}{2} 
using the log$\N{Si^+} = 16.0$ measurement from HIRES as reference.  
In the HIRES spectrum, we measure the \ion{B}{2} $W_{obs} \approx 18$ m\AA\ 
for a SNR $\approx 30$ (per 2 \kms pixel).
The offset between the observed and predicted $W_{obs}$ values is due to our assumption 
that the column densities scale with their solar abundances with no corrections.  
Indeed, in this case the gas-phase 
B/Si abundance is super-solar and we therefore expect the 
observed value to be found above the predicted value.  

One can then determine how the velocity width V of a line relates to SNR
for a fixed $W_{obs}$.  Lines near $W_{obs}=10$ m\AA\ should be trivial to detect 
at $3\sigma$ with HIRES.
Lines below $W_{obs}=5$ m\AA\ quickly become too difficult 
to detect (for $W_{obs}=5$ m\AA\ and V $=$ 4, 10 and 20 km s$^{-1}$, SNRs 
are $\approx$ 25, 40, and 55, respectively) and features 
near $W_{obs}=1$ m\AA\ are at the HIRES detection 
limit for faint quasars (for V $=$ 4 km s$^{-1}$, SNR is $> 100$; 
for V $=$ 16 km s$^{-1}$, SNR is $> 200$).

Therefore, we determine that lines with $W_{obs} \ge 5$~m\AA\ would 
be detectable with observations similar to those of DLA-B/FJ0812+32. 
This implies that many \ion{B}{2} 1362 and \ion{O}{1} 1355 lines, if in detectable 
regions of the spectrum, will be observed with future HIRES 
observations.  Many of the predicted Sn, Pb, and Kr equivalent widths are 
currently out of reach (for example 
$W_{obs} = 2.5$, 1.8 and 4.3 m\AA\, respectively, for DLA-B/FJ0812+32 
assuming log$\N{Si^+} = 16.0$), but our 
SDSS metal-strong sample includes a number of excellent candidates for
observations with future instruments.

\section{Implications for Damped \lya\ Systems}

\subsection{Metal-Strong DLAs}

Given the results from our ESI observations, we 
may now return to our SDSS MSDLA candidates and
estimate the fraction of DLAs that are truly metal-strong.  This may help 
us determine if 
previous analyses of DLA chemical evolution have suffered 
from small sample sizes.  Our SDSS-DR3 DLA Survey (PHW05) 
presents $\mnhi$ measurements 
of 525 DLAs automatically detected 
in the SDSS-DR3 QSO sample, which is $\approx 10\times$ larger than the 
combined QSO sample size of previous samples 
at $z_{abs}\approx 3$ \citep{peroux03}.  
Noting that the SDSS-DR3 DLA Survey is 
$>95\%$ complete (and $100\%$ for log $\mnhi > 20.4$, as most 
metal-strong DLAs tend to be), we may now examine the overall incidence of 
MSDLAs. 

We find $\frac{29}{304} \approx 10\%$ of all DLAs in our SDSS-DR3 DLA Survey (restricted to $r < 19.5$) 
to be candidate S-DLA absorbers, and 
$\frac{15}{304} \approx 5\%$ 
of all DLAs in the same sample 
to be candidate VS-DLA absorbers.  Recall that these ratings 
are largely qualitative and based on the observed absorption depths 
of \ion{Si}{2} 1808 profiles in the SDSS spectra (see $\S$~3.1).  We present these separate categories here to compliment Table~\ref{all_DR3} and so that the 
reader may appreciate the differences between the candidate samples.  
In our ESI sample of 27 QSOs, 
$\frac{3}{10} = 30\%$ of our S candidate absorbers are 
confirmed as truly metal-strong 
(i.e.\ log$\N{Zn^+} \ge 13.0$ or log$\N{Si^+} \ge 15.8$ in ESI data, 
accounting for possible saturation effects) 
while $\frac{9}{17} \approx 53\%$ of 
our VS candidate absorbers are truly metal-strong.  
We do not include the upper limits here, 
even if they lie above the metal-strong threshold.  
We therefore propose that $(0.30 \times 29 + 0.53 \times 15)/304 \approx 5\%$ of all DLAs 
with $z_{abs} \ge 2.2$ observed in QSO sightlines with $r < 19.5$ are truly metal strong.  
This agrees with a similar estimate ($< 9\%$) for metal-strong systems as determined from 
simulations (Ellison 2005 and references therein).
Such a result helps explain why previous 
studies 
have never clearly identified the MSDLA population, as sample sizes had 
not been large enough until now.

\subsection{Dust Obscuration Bias?} 

As described in $\S$~2, we define metal-strong absorbers to have 
log$\N{Zn^+} \ge 13.15$ in accordance with DLA-B/FJ0812+32, 
also corresponding to the \cite{boisse98} obscuration 
`threshold'.  
However, as it may have been misunderstood in previous works, 
we emphasize that the Boiss\'e threshold was not 
intended to be a defining boundary of a `forbidden region' of 
DLA absorbers (P.\ Boiss\'e, private communication).  The author 
maintains that a small percentage of DLAs 
are expected to lie beyond this threshold and should 
be revealed as optical samples begin probing fainter QSO sightlines.
We will argue that at least the region near log$\N{Zn^+} = 13.15$
is not disfavored by a statistically 
significant 
dust bias but that these systems are simply rare.
 
\begin{figure}[ht]
\centerline{
\includegraphics[angle=90,width=3.4in]{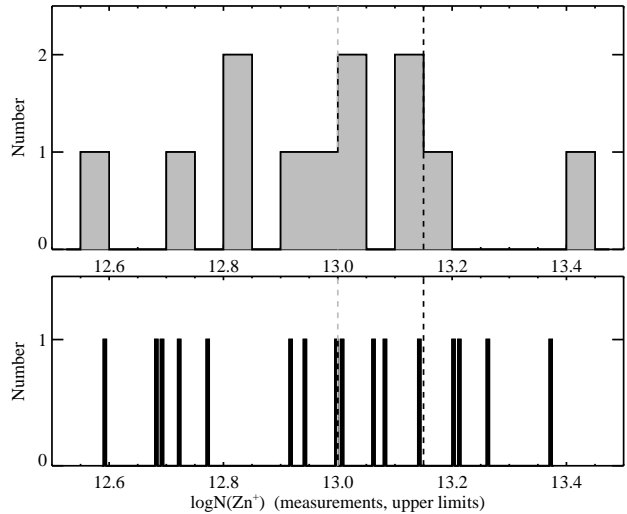}}
\caption{Histogram of log$\N{Zn^+}$ from the ESI data (including DLA-B/FJ0812+32, 
with log$\N{ZnII\ 2026} = 13.04 \pm 0.02$).
Measurements are plotted as filled gray in the top panel, 
upper limits as black lines in the bottom panel,
and the metal-strong 
threshold (log$\N{Zn^+} \approx 13.15$) as a dashed black line.  
Note that systems with log$\N{Zn^+} \approx 13$ (dashed gray line) 
may be underestimated due to line saturation in ESI spectra.
\label{fig:zn_hist}}
\end{figure}

Figure~\ref{fig:zn_hist} presents a histogram of 
log$\N{Zn^+}$ from our ESI data.  Measurements are plotted 
as filled gray in the top panel, 
upper limits as black lines in the bottom panel, and 
the metal-strong threshold (log$\N{Zn^+} = 13.15$) as a dashed black line.
We emphasize that ESI measurements of Zn lines near log$\N{Zn^+} \approx 13.0$ 
(dashed gray line) are 
systematically underestimated by 0.1 to 0.2 dex due to line saturation.  
Therefore, nearly half of the detections in Figure~\ref{fig:zn_hist} probably 
match or exceed the metal-strong threshold.

\begin{figure}[ht]
\centerline{
\includegraphics[angle=90,width=3.4in]{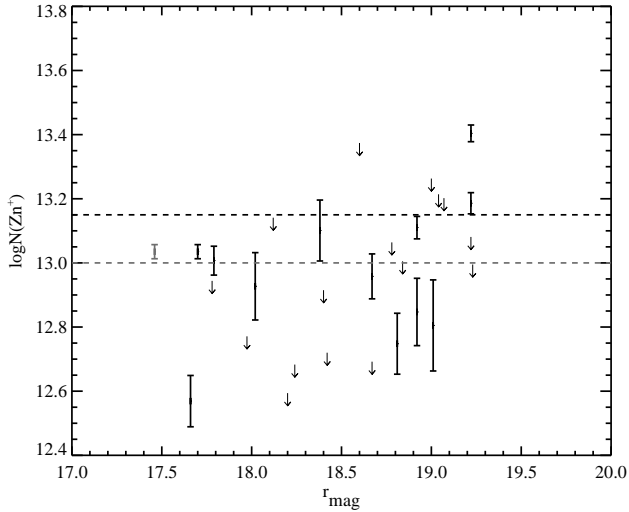}}
\caption{log$\N{Zn^+}$ vs.\ $r$ magnitude from our ESI sample.  
The metal-strong 
threshold, log$\N{Zn^+} = 13.15$, is marked as a dashed black line.  
DLA-B/FJ0812+32 is marked in gray.  
Note that systems with log$\N{Zn^+} \approx 13$ (dashed gray line) 
may be 
underestimated due to line saturation in ESI spectra.
A nearly 2 magnitude statistically constant 
scatter in the magnitudes of 
metal-strong sightlines is observed; a Spearman rank correlation test 
on the measured values (excluding limits) gives a linear correlation 
coefficient of 0.49 at $<2\sigma$ statistical significance.  Finding MSDLAs along fainter 
QSO sightlines (ie. $r > 19.5$) will provide a better means to test the effects 
of dust on this population.
\label{fig:zn_mag}} 
\end{figure}

As mentioned in \cite{boisse98} and originally developed in \cite{fall93}, 
significant dust extinction might preferrentially 
select QSO systems having a low metal content.  This follows from the argument that 
metal-strong systems having a high metal content would likely also have significant 
dust columns, thereby considerably dimming the 
background QSO.  In magnitude-limited surveys, 
therefore, one might expect to find the most 
metal-strong absorbers towards the faint end 
of the QSO-sightline magnitude distribution.  
Figure~\ref{fig:zn_mag} presents log$\N{Zn^+}$ vs.\ $r$ magnitude from our ESI sample.
Keeping in mind that systems near log$\N{Zn^+} \approx 13.0$ 
(dashed gray line) are likely underestimated 
due to saturation, one notices a significant scatter in the $r$ 
magnitudes consistent with no statistically significant dust bias; a 
Spearman rank correlation test 
on the measured values (ie.\ excluding limits) gives a linear correlation 
coefficient of 0.49 at $<2\sigma$ statistical significance.   
However, we emphasize 
that the true test of this debate lies in observing fainter QSOs.  
Indeed, our strongest Zn absorber (SDSS1610+4724) does lie among 
the faintest QSOs in our metal-strong sample; 
this system in particular is most likely affected by dust obscuration 
(the results presented in $\S$~8 find it to have the highest [Mn/Fe] value 
in the sample, $+0.2 \pm 0.1$ dex, as well as high [Zn/Fe] ,$0.7 \pm 0.1$ dex, 
both telltale signatures of dust).  And, we also 
observe a lack of clearly metal-strong 
systems along the brightest QSO sightlines.  

The \cite{boisse98} sample contained 
37 DLAs, and considering the MSDLA fraction is $\approx 5\%$ it is not 
surprising that the number of MSDLAs described in that work was zero (also note that 
the limiting magnitude there was significantly brighter than ours, near $r \approx 
18.5$, so systems predicted to be the most strongly affected by dust, the MSDLAs, were 
even more unlikely to be found in that sample).  We contend, however, 
that a significant dust bias is not evident in either our DLA or MS absorber samples 
and that MSDLAs are not often discovered at high redshift 
simply because they are rare.  
However, we acknowledge 
the hindrance of not having HI measurements for most of these systems; 
we have therefore not computed dust-to-gas ratios or ionization 
fractions for the absorbers in our sample.  
These values would be critical for a thorough analysis of 
dust extinction in metal-strong absorbers, and we plan to confront 
such issues in future papers.

Other works have also been unable to find evidence supporting a significant 
dust bias on the overall DLA population.
Murphy \& Liske (2004) presented a spectrophotometric survey of 
over 70 DLA sightlines from SDSS-DR2
comparing the 
spectral index distribution of these DLA sightlines with a large control sample and 
found no evidence for dust-reddening at $z \approx 3$.  They 
placed a limit on the shift of 
the spectral index $|\Delta \alpha| < 0.19$ ($3 \sigma$), corresponding to 
$E(B-V) < 0.02$ mag ($3 \sigma$) for SMC-like dust, and 
preliminary SDSS-DR3 results 
show $\Delta \alpha = -0.063 \pm 0.014$ ($1 \sigma$ error; $\Delta \alpha$ 
significant at $\approx 4.5 \sigma$), 
corresponding to $E(B-V) = 0.0071 \pm 0.0016$ mag and 
$A_{v} \approx 0.02$ mag for SMC-like dust 
(Murphy et al.\ 2006; in prep.).  
The $E(B-V)$ value is derived by (de-)reddening the DLA QSO spectra according to  
the SMC-like dust law with different values of $E(B-V)$ in a maximum likelihood analysis.
The authors caution, however, that this preliminary result does not yet include 
any correction for the color selection of the QSOs in SDSS.  They expect their $E(B-V)$ result 
to be somewhat more positive and significant once this is taken into 
account, with a very preliminary, rough estimate of $\approx < 40\%$ effect.  This 
comes from estimates of SDSS completeness at different values of the spectral index in 
Murphy \& Liske (2004).  Although the nature of dust in high-redshift 
galaxies is not well understood, these results indicate a very 
small reddening of the DLA QSO sightlines and 
that dust obscuration is therefore not very important for 
the overall DLA population.  The results are inconsistent with earlier studies of 
Fall, Pei and collaborators, which the authors attribute to the small 
sample sizes used in previous works.  

Furthermore, 
CORALS I \citep{ellison01} demonstrated that $z > 2$ samples of DLAs toward radio-selected 
quasars show no significant difference from optically-selected samples, 
and CORALS II \citep{ellison04} indicates 
only a mild effect at lower redshifts where integrated star formation histories 
are most significant.   \cite{ellison05} finds $E(B-V) < 0.04$ in CORALS DLAs when 
comparing optical-to-infrared colors of QSOs with and without intervening absorbers.  Finally, \cite{akerman05} also found 
no evidence for increased dust depletions in CORALS DLAs and 
stated that large scale optical QSO surveys give a fair census of the high-redshift absorber population.  However, 
it should be noted that the CORALS results come from small sample sizes and may not 
be fully representative of the overall DLA population.

If a significant dust bias did exist 
one would expect to observe higher 
\nhi\ values towards fainter QSOs, as these systems would likely have a higher 
dust content.
The SDSS-DR3 DLA Survey (PHW05) presents results that we 
argue are contradictory to this idea.  
That work (which includes 525 SDSS DLAs with $z_{abs}>2.2$) shows 
higher \nhi\ systems towards {\it{brighter}} QSOs.  In the paper, PHW05 
discuss a variety of systematic 
effects that may be responsible and conclude with the possibility of 
gravitational lensing (GL).  Indeed, \cite{murph04} measured an 
$\approx 2 \sigma$ excess of bright and/or deficit of faint SDSS-DR2 
QSOs with intervening DLAs and attributed this to GL.  
That group is currently examining this effect with the increased 
SDSS-DR3 DLA sample.
However, we caution that the effects of dust and GL 
are quite difficult to disentangle and 
prefer to refrain from further comment until the interplay between 
these competing effects is better understood.

In the interest of a balanced discussion of the dust bias on DLAs, 
we wish to mention recent works supporting the dust bias and touching 
on strong metal-absorption systems.  Notably, \cite{vladilo05} derived a 
relation between the extinction of a DLA system 
and its $\mnhi$ value, metallicity $Z$, fraction of iron in dust, $f_{\rm{Fe}}(Z)$, and 
redshift $z_{abs}$.  The authors argued that this relation predicts that 
$\approx 30 - 50\%$ of all DLAs are missed as a result of their own extinction 
in magnitude-limited surveys, and show that the empirical thresholds of 
log$\N{Zn^+} \approx 13.2$ and log$\mnhi \approx 22$ are also quantitative predictions of 
their model.  We claim that such systems are simply rare and therefore not often 
discovered.  Of the handful of MSDLAs we do find in a quasar set of nearly 20,000, 
we make special note of SDSS1610+4724 with log$\N{Zn^+} =  13.40 \pm 0.03$, clearly 
above the proposed extinction `threshold' (yet admittedly in a faint QSO sightline, 
$r = 19.22$).  The author has proposed that the extinction per metal column density
might drop in interstellar environments with extremely high density owing to 
coalescence of dust grains, and that SDSS1610+4724 
could be one of those rare, very interesting cases  
(G.\ Vladilo, private communication).  If this is not the case, however, 
SDSS1610+4724 may pose a (perhaps minor) 
challenge to their model of dust obscuration.  
Unfortunately, the current lack of Zn measurements precludes 
an exact estimate of the bias;  this makes MSDLAs the 
critical subclass for gauging the effect.  

Also important to these issues is the resemblance between 
the properties of local dust grains and those in 
high-redshift clouds;  \cite{vladilo06} have recently investigated this issue 
in absorbers out to 
$z \approx 2$ and find that the
mean extinction per atom of iron in the dust is remarkably similar to that found
in interstellar clouds of the Milky Way.  Also noted in their paper is the previous study by 
\cite{petitjean02} of SDSS0016--0012, a system also found in our sample and with a velocity profile 
spanning roughly 1,000 \kms .  \cite{petitjean02} found SDSS0016--0012 
to have a high dust content, as well as 
the highest overall (${\rm{H}}_{2}$) molecular fraction of DLAs at that time, and argued in 
support of a dust obscuration bias.

We also comment on the large, recent survey of \cite{york06}. 
This group examined 809 \ion{Mg}{2} absorption systems 
in SDSS and found that the average 
extinction curves of their sub-samples are similar to the 
SMC extinction curve with a rising UV extinction below 2200\AA. The authors also 
found that the 
absorber rest frame color excess, $E(B-V )$, derived from the extinction curves, depends on
the absorber properties and ranges from $<0.001$ to 0.085 for various sub-samples.  While a 
notable result, we argue that even systems with $E(B-V) \approx 0.1$ are 
unlikely to provide a statistically significant 
dust obscuration bias on the {\it{overall}} DLA population. 

Some highly dusty systems have also been discovered at 
lower redshifts ($z \approx 1$) by \cite{wild_05} and \cite{wild_06}, via strong 
\ion{Ca}{2} absorption.  Note that these systems 
are inferred DLAs from their corresponding \ion{Mg}{2}, \ion{Mg}{1} and \ion{Fe}{2} absorption features.  A composite spectrum from the \cite{wild_05} sample yields $E(B-V) \approx 0.06$, 
while the absorbers from \cite{wild_06} have on average $E(B-V) > \approx 0.1$.

Finally, to add an overlying word of caution 
to this entire debate, note the study of 
\cite{hopkins04} who examined 
SDSS quasars and found that reddening along these 
lines of sight is dominated by SMC-like dust {\it{at the quasar redshifts}}; that is, 
not even primarily due to intervening absorbers.

\subsection{Effects on the $\mnhi$-weighted Cosmic Mean Metallicity $<Z(z)>$}

\cite{pro03a} demonstrated a statistically significant 
evolution in the $\mnhi$-weighted cosmic 
mean metallicity $<Z(z)>$ from $z_{abs} > 2$ DLA absorbers. 
Here we will investigate how MSDLAs could influence their measurements.  
The chemical-enrichment (C-E) sample of \cite{pro03a} included 
113 DLAs with $z_{abs} \ge 1.6$ and in QSO sightlines with $r < 19.5$\,mag 
and 95 DLAs with $z_{abs} \ge 2.2$.
When considering the $z_{abs} \ge 1.6$ C-E sample, 
we find an evolution in $<Z(z)>$ 
to be $m = -0.27 \pm 0.03$, consistent with the \cite{pro03a} value (not 
cut for $z_{abs} \ge 1.6$).  
Recall that $\approx 5\%$ of all $z_{abs} \ge 2.2$ 
DLAs in QSO sightlines with $r < 19.5$ mag 
are expected to be metal-strong.  
We may then roughly estimate the number of 
MSDLAs expected in a given DLA sample; 
we might expect to find roughly five MSDLAs in the $z_{abs} \ge 2.2$ 
C-E DLA sample.  That sample currently contains only 
one such system, DLA-B/FJ0812+32, the defining MSDLA 
(recall that log$\N{Zn^+} = 13.04$ from ESI, whereas HIRES data shows 
log$\N{Zn^+} = 13.15$; an example of the above-mentioned ESI saturation issue).

\begin{figure}[ht]
\centerline{
\includegraphics[angle=90,width=3.4in]{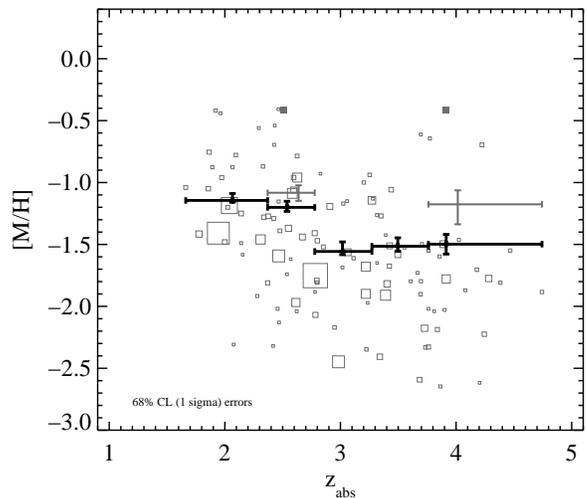}}
\caption{$\mnhi$-weighted cosmic mean metallicity $<Z(z)>$.  
DLAs are shown as squares scaled to the $\mnhi$ of the system 
(open are from the CE sample, filled are the added SDSS1610+4724
MSDLA).  $<Z(z)>$ is plotted for 5 bins with 
$1\sigma$ uncertainties given by bootstrap analysis.  
The dark gray errors for the 
$z_{abs} \approx 2.6$ \& 4 bins are the 
new $<Z>$ values in each bin (offset by 0.1 in $z$ for presentation) 
after one MSDLA, SDSS1610+4724, is added.
\label{CE}}
\end{figure}

To estimate the effect {\it{of a single}} MSDLA, we add 
\\SDSS1610+4724 (our strongest absorber in 
Zn, log$\N{Zn^+} = 13.40 \pm 0.03$, and with measured 
log$\N{HI} = 21.15 \pm 0.15$; [Zn/H]$=-0.42 \pm 0.15$)
to the 
C-E sample.
We justify this exercise by the reasoning stated above 
(ie.\ expecting $\approx 5$ MSDLAs in this sample), 
yet we note that adding many more of these systems without also including 
their corresponding non-MSDLAs would bias the C-E sample to an 
unjustifiable extent. 
Figure~\ref{CE} illustrates that adding this one 
MSDLA raises $<Z>$ in the $z \approx 2.6$ bin 
by +0.12 dex, ie. $\approx 2\sigma$.  
Including this system does not significantly change $m$, however, just a 
slight increase in the scatter: now $m = -0.27 \pm 0.04$\,dex.

Instead, how might including this same metal-strong 
absorber at a higher redshift affect $<Z(z)>$ and its evolution?  
Adding SDSS1610+4724 to the $z \approx 4$ bin of the C-E sample raises $<Z>$ 
from -1.50 to -1.18 (ie.\ a +0.32 dex 
or $\approx 2\sigma$ increase, yet 
note that such an object being found there is highly 
unlikely given the current size and statistics of the $z \approx 4$ bin).
Nevertheless, the linear $<Z(z)>$ evolution slope 
also remains nearly identical in this case, $m = -0.26 \pm 0.04$, driven 
by the data with smaller errors.  
We therefore suggest that 
MSDLAs could have a modest impact on the $<Z>$ of a single bin yet 
little impact on the overall DLA $<Z(z)>$ evolution.  
However, in the unlikely event that one MSDLA 
was found in each bin under similar statistics, 
the effect on the linear $<Z(z)>$ evolution slope could be larger.

\section{Abundance Analysis from the ESI Sample}

We now examine certain abundance ratios from our ESI data and comment on 
the dust depletion vs.\ SNe enrichment `debate'.  
Although HIRES data is superior, ESI can still be used to examine global abundance trends.  

Before we begin, we must mention here that one system, SDSS0927+5621 ($z_{abs} = 1.78$), has a confirmed 
log$\mnhi = 19.00^{+0.10}_{-0.25}$ (Prochaska et al.\ 2006, submitted).
This system was selected 
by its metal-strong signature, and metal-strong systems usually are in
the $\mnhi$ range of DLAs.  Why is this object both \ion{H}{1}-poor and metal-strong?
Prochaska et al.\ (2006, submitted) use HIRES data and find that SDSS0927+5621 is 
highly ionized, with 
an ionization factor $x \approx  0.9$ (based on inferences made from the observed 
$\mnhi$ and ${\rm{Al}}^{++}$ values).  
In this case, the QAL system is too far from the 
QSO to be ionized by it ($z_{em}=2.28$).  
The extragalactic UV background (EUVB; from QSOs and/or galaxies not in the sightline) is 
surely a part of the responsible ionizing radiation, 
but the dominant component likely comes from young, massive stars within the host galaxy.  
If massive stars are present, one might expect strong-metal absorption (observed), 
assuming some previous enrichment and energetic feedback processes.  Indeed, the authors 
find SDSS0927+5621 to have wide, complex velocity profiles and propose that such 
kinematic structure is indicative of feedback processes correlated with star formation.  
The authors also (tentatively) find SDSS0927+5621 to have the highest gas metallicity of any 
astrophysical environment and total metal surface density exceeding nearly every known DLA.  
This system clearly falls into a separate category of QAL systems, the (super-solar) super-LLS, 
and so we remove it from the following 
abundance analysis.  We caution the reader that the possibility 
of other assumed DLAs (with $z_{abs}<2.2$) 
falling into this `ionized' category remains.  However, recall that 
only $\approx 1\%$ of our candidate MS SDSS sample with $z_{abs} \ge 2.2$ was 
below the DLA threshold.  
We await \nhi\ measurements from upcoming observations 
to further comment on this issue.
  
We also require the systems to have Fe measurements 
due to the important role of Fe in abundance ratios; therefore, 
SDSS0316+0040 and SDSS1235+0017 are also 
excluded from the following analysis.  
Also note that the relative abundances to 
be plotted will not involve \nhi.\footnote{For example,  
\begin{equation}
[{\rm Si}/{\rm Ti}]=[{\rm Si}/{\rm H}]-[{\rm Ti}/{\rm H}]={\rm log}\N{Si}-{\rm log}\N{Ti}-{\rm log}\N{Si}_{\sun}+{\rm log}\N{Ti}_{\sun} \;\;, 
\label{eqn:abndex}
\end{equation}
\noindent
and therefore the relative abundances are independent of 
$\mnhi$.}
We have assumed a conservative lower limit of $0.10$\,dex 
error ($1\sigma$) for all measurements 
and emphasize that such large errors are the most significant hindrance to doing 
abundance-ratio analysis with this moderate-resolution data.

\begin{figure}[ht]
\centerline{
\includegraphics[angle=90,width=3.4in]{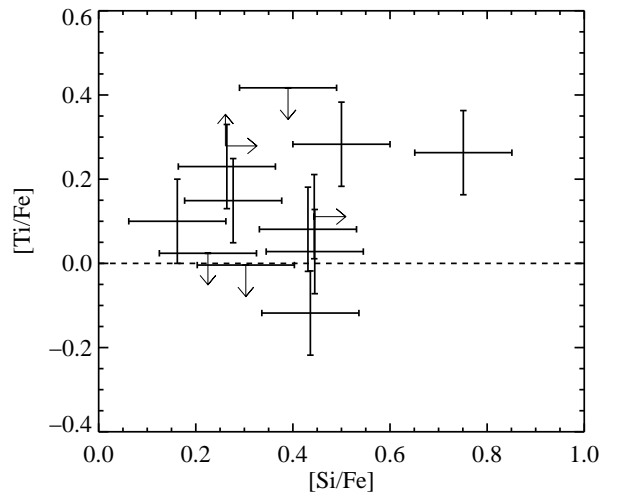}}
\caption{[Ti/Fe] vs.\ [Si/Fe] from our ESI sample.  The enhancement of
Ti/Fe suggests the gas is predominantly enriched by Type~II SNe.
\label{TiFe}}
\end{figure}

Figure~\ref{TiFe} shows a plot of [Ti/Fe] vs.\ [Si/Fe]  
\footnote{Note that the outlying [Ti/Fe] measurement of 1.14 (the plot is 
cropped and so does not show this point) is from SDSS0016--0012, 
a system which displays particularly wide absorption profiles, some spanning 
roughly 1,000 \kms.  We do not place much confidence in 
measurements resulting from integrating shallow absorption across such
wide profiles.}.
It is evident here that all the 
systems in our sample show enhanced Si/Fe ratios.  This can 
be interpreted as either Type II SNe enrichment or 
depletion of Fe onto dust, or both.  
Also note that at low [Si/Fe] the [Ti/Fe] values are enhanced.  
Although Ti behaves like a refractory iron-peak element, in Galactic stars it 
shows a similar trend as the $\alpha$ elements and is thus generally accepted 
as an $\alpha$ element; therefore, because both Ti and Fe are refractory and 
because Ti is more heavily depleted than Fe \citep{savage96}, 
an overabundance of [Ti/Fe] at super-solar [Si/Fe] 
implies nucleosynthetic $\alpha$-enrichment \citep{miro02b}.  
An underabundance of [Ti/Fe], however, may be attributed to depletion of gas 
onto dust, and we present a few of these candidates here as well.  Overall, 
this plot may suggest Type II SNe enrichment in many of the MS systems 
presented here.

\begin{figure}[ht]
\centerline{
\includegraphics[angle=90,width=3.4in]{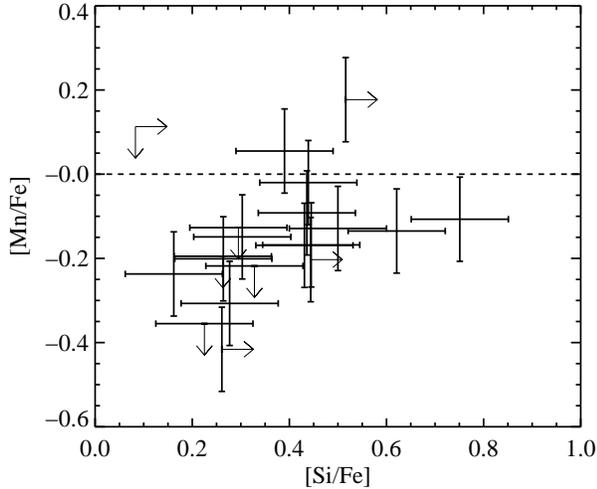}}
\caption{[Mn/Fe] vs.\ [Si/Fe] from our ESI sample.  
Systems with [Mn/Fe]~$> 0$ and [Si/Fe]~$> +0.3$ must 
be strongly affected by depletion of gas onto dust.
\label{MnFe}}
\end{figure}

Figure~\ref{MnFe} shows a plot of [Mn/Fe] vs.\ [Si/Fe]. 
The [Mn/Fe] ratios are mostly sub-solar, 
and agree with previous measured values attributed to nucleosynthetic enrichment 
\citep{lu96,ledoux98}  including from Type Ia SNe \citep{miro02b}.
We expect these observations to support metallicity-dependent 
Mn yields from both Type Ia and Type II 
SNe (McWilliam, Rich, \& Smecker-Hane 2003), 
but require log$\mnhi$ measurements to be certain.  
In a future paper, we will investigate the differences in Mn yields from Type Ia- 
and Type II SNe-dominated systems; recent models indicate that 
particular distinction may soon 
be discovered at high [Mn/Fe] values.  We also aim to address the 
range of dust-depletion levels in the MSDLAs.  
Here we present a new super-solar  
[Mn/Fe] value and another just shy due to the error; 
these systems present strong evidence for substantial
dust depletion. 

\begin{figure}[ht]
\centerline{
\includegraphics[angle=90,width=3.4in]{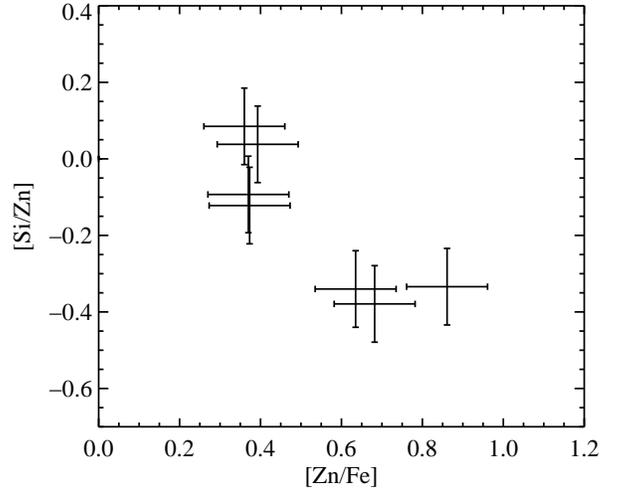}}
\caption{[Si/Zn] vs.\ [Zn/Fe] from our ESI sample, for all systems with 
measured values of each element.  
Systems with low [Si/Zn] and high [Zn/Fe] are the strongest affected by depletion 
effects.
\label{SiZnFe}}
\end{figure}

\begin{figure}[ht]
\centerline{
\includegraphics[angle=90,width=3.4in]{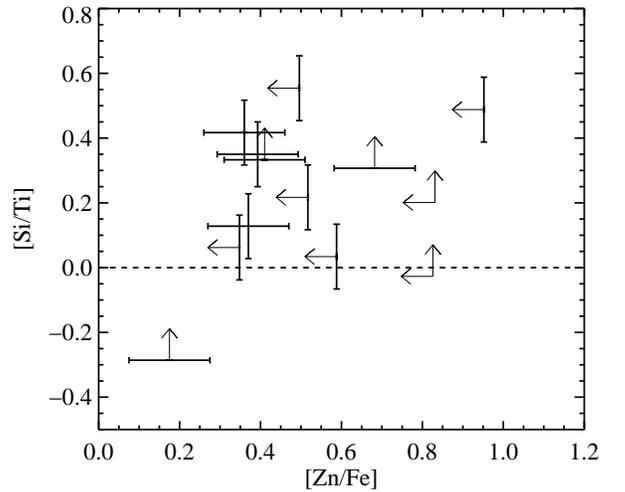}}
\caption{[Si/Ti] vs.\ [Zn/Fe] from our ESI sample.  This plot supports 
depletion of gas onto dust grains.
\label{SiTi}}
\end{figure}

Figure~\ref{SiZnFe} shows [Si/Zn] vs.\ [Zn/Fe] for systems 
in our sample with both Si and Zn measurements.
One would expect 
systems displaying subsolar [Si/Zn] values to be more affected by dust 
depletion than by nucleosynthetic effects because Si 
is more refractory than Zn.  
What is observed suggests that some 
MS systems may indeed contain high amounts of dust 
(points at high [Zn/Fe] and low [Si/Zn]). 
Also supporting this idea is Figure~\ref{SiTi}, where we plot [Si/Ti] 
vs.\ [Zn/Fe].  
Because Ti is the most refractive, Si/Ti enhancements at super-solar Zn/Fe are likely 
due to the depletion of Ti and Fe onto dust grains.  \cite{pw02} state 
that a correlation between these ratios is strong evidence for dust depletion, and although 
there is a large scatter the limits suggest a weak overall correlation.  
While some MSDLAs likely contain a large amount 
of dust and may eventually be observed to have 
their own dust-obscuration bias 
(MSDLA samples toward fainter QSOs will best address this issue), 
we maintain that the {\it{overall}} DLA population is not significantly 
affected by a dust bias (see previous sections). 

\section{Summary and Conclusions}

To confirm that our automated technique works well, we 
summarize our metal-strong detections as follows.  
Our algorithms searched 19,429 SDSS-DR3 QSO 
sightlines with $z_{QSO} \ge 1.6$ and found 435 
visually-confirmed candidate metal-strong absorbers.
Of these metal-strong candidate systems 
with \lya\ wavelength coverage available in the SDSS, 
we find $< 5\%$ (and likely near $\approx 1\%$) are
without corresponding DLAs, ie.\ systems belonging to the super-LLS population.  

Our metal-searching technique is most sensitive 
to systems with large $W$ metal profiles, but recall that a large $W$ is 
not a reliable predictor of $N$ for a saturated transition.  
Therefore, we focused on weak lines as metal-strong 
indicators.  We plan to use these MS indicators to determine which systems 
may be directly observed at high resolution without first obtaining medium-resolution 
confirmation spectra.  We argued that $W_r(\rm ZnII\ 2026)$ is the best 
indicator of metal-strong QAL systems and have compared our 
sample to the SDSS-DR2 spectrum of the metal-strong 
DLA-B/FJ0812+32 DLA.  We also claimed 
that $W_r(\rm SiII\ 1808)$ is not as reliable an indicator 
as \ion{Zn}{2} 2026 but that it can still be useful for faint QSO sightlines.  

All of our ESI (\ndla\ metal-strong systems with $r < 19.5$ mag) 
sample had the highest score possible from 
our SDSS-searching algorithms.  
Our ESI sample is a collection of 10 and 17 
subjectively-rated S and VS systems in the SDSS, respectively.
Finally, we have measured (with ESI) 11 systems with $\N{Si^+}$ values 
near or greater than DLA-B/FJ0812+32 (as measured from ESI data) and 5 with
$\N{Zn^+}$ greater than or equal to the Zn detection from 
DLA-B/FJ0812+32 (in the ESI data -- recall the importance of 
saturation effects in MSDLA classification; the HIRES spectrum of 
DLA-B/FJ0812+32 yields log$\N{Zn^+} 
\\\approx 13.15$).  
Therefore, $\approx 40\%$ of the systems in our ESI sample yielded $\N{Si^+}$ or 
$\N{Zn^+}$ measurements consistent with or higher than DLA-B/FJ0812+32.
PHW05 demonstrated that our automated SDSS DLA-finding technique 
also works well, and we will search future 
SDSS Data Releases for MSDLAs and publish updated 
metal-strong candidate lists periodically.

We estimated the feasibility of detecting certain weak transitions at $3\sigma$, 
predicting what SNRs we would need to reach with HIRES-grade instruments.  
We find that lines with 
$W_{obs} \ge 5$ m\AA\ would be detectable with observations similar to DLA-B/FJ0812+32, 
and therefore that a handful of \ion{B}{2} 1362 and \ion{O}{1} 1335 lines may 
soon be observed in our metal-strong ESI sample.  
As the collecting power of modern and future telescopes continues to increase 
(and with it SNRs), we will begin detecting Sn, Pb, and Kr lines in these 
high-redshift systems.  
We are currently testing our predictions 
with KeckI/HIRES  and VLT/UVES observations.

Taking our SDSS and ESI results 
together, we predict that $\approx 5\%$
of all $z_{abs} \ge 2.2$ DLAs in QSO sightlines with $r < 19.5$ 
are truly metal-strong.  This result is of 
particular importance when considering previous results from DLA 
studies with small sample sizes, especially 
those proposing a significant dust-obscuration bias on the overall 
DLA population.  Along separate lines of research 
and with additional results from our SDSS-DR3 DLA Survey, 
we find no evidence in support of a statistically 
significant dust-obscuration bias for the overall DLA population.  
We await a larger, deeper 
metal-strong DLA sample, complete with UV \lya\ 
measurements to further comment on this issue and its 
relevance to the MSDLA population.

We then investigated how a single MSDLA might 
affect the $\mnhi$-weighted cosmic mean metallicity $<Z(z)>$ as 
determined from previous DLA studies.  
We find that adding our strongest Zn absorber (SDSS1610+4724) to the 
current C-E sample raises $<Z>$ near $z \approx 2.6$ by +0.12 dex 
($\approx 2\sigma$).  
When the same MSDLA is instead added at higher 
redshift (albeit a more extreme and perhaps currently unrealistic proposition), 
we find a +0.32 dex increase ($\approx 2\sigma$) in the $z \approx 4$ bin.
The overall linear slope of the $<Z(z)>$ evolution remains 
essentially unchanged in both cases. 
We therefore contend that MSDLAs may be significant when 
considering $<Z>$ in a particular bin, yet given the current 
statistics are not a strong influence on the 
evolution of  $<Z(z)>$ from the overall DLA population.

Although the conservative errors assumed for the relative abundances from 
our Keck ESI (medium-resolution) data are 
large, and we lack many \lya\ measurements for 
computing ionization fractions and dust-to-gas ratios, 
we concluded with a brief discussion of 
abundance ratios from our ESI sample and   
find evidence for significant dust depletion 
in a handful of systems underlying largely Type II SNe enrichment.  

\section{Acknowledgements}

We thank the SDSS team for their incredible survey and the Keck staff for their assistance and 
hospitality.  We acknowledge the privilege of observing from the summit of Mauna Kea as it 
has long been considered a sacred site within the indigenous Hawaiian community.  
We thank S.\ Burles for providing the SDSS continuum-fitting PCA software.  
Thanks to P.\ Boiss\'e, G.\ Vladilo, D.\ York, J.\ Bergeron, and N. Prantzos for helpful comments and 
suggestions, and to M.\ Murphy for kindly providing and discussing his SDSS-DR3 DLA-reddening 
results prior to publication.  We also thank the referee for 
helpful suggestions.  SHF especially thanks the UCSC Department of Physics for 
the generous Undergraduate Thesis Award that helped make 
his December 2003 travel to Keck possible.  
JXP and SHF were partially supported by NSF 
grant AST-0307408 and its REU sub-contract.
JXP and AMW are partially supported by NSF grant
AST-0307824.  



\clearpage

\begin{deluxetable}{lccc}
\tablewidth{300pt}
\tablecaption{Atomic Data \label{tab:fosc}}
\tabletypesize{\footnotesize}
\tablehead{\colhead{Transition} &\colhead{$\lambda$ (\AA)} &\colhead{$f$} & \colhead{Ref.}}
\startdata
 \ion{H}{1}-A 1215 & 1215.6701 & 0.4164 &  1  \\
  \ion{Kr}{1} 1235 & 1235.8380 & 0.1871 &  1  \\
 \ion{Si}{2} 1260$^a$ & 1260.4221 & 1.007 &  1 \\
   \ion{O}{1} 1302$^a$ & 1302.1685 & 0.04887 &  1 \\
 \ion{Si}{2} 1304$^a$ & 1304.3702 & 0.094 &  2 \\
 \ion{Ni}{2} 1317 & 1317.2170 & 0.058 &  14  \\
  \ion{C}{2} 1334$^a$ & 1334.5323 & 0.1278 &  1 \\
 \ion{C}{2*} 1335 & 1335.7077 & 0.1149 &  1  \\
   \ion{O}{1} 1355 & 1355.5977 & 1.25E-6 &  1  \\
  \ion{B}{2} 1362 & 1362.4610 & 0.987 &  1  \\
 \ion{Ni}{2} 1370 & 1370.1310 & 0.0769 &  3  \\
 \ion{Si}{4} 1393 & 1393.7550 & 0.528 &  1  \\
 \ion{Sn}{2} 1400 & 1400.4000 & 1.0274 & 12  \\
 \ion{Si}{4} 1402 & 1402.7700 & 0.262 &  1  \\
 \ion{Pb}{2} 1433 & 1433.9056 & 0.87 &  1  \\
 \ion{Ni}{2} 1454 & 1454.8420 & 0.0323 &  4  \\
 \ion{Ni}{2} 1467 & 1467.2590 & 6.3E-3 &  4  \\
 \ion{Ni}{2} 1467 & 1467.7560 & 9.9E-3 &  4  \\
 \ion{Si}{2} 1526$^a$ & 1526.7066 & 0.127 &  5 \\
  \ion{C}{4} 1548 & 1548.1950 & 0.1908 &  1  \\
  \ion{C}{4} 1550 & 1550.7700 & 0.09522 &  1  \\
   \ion{C}{1} 1560 & 1560.3092 & 0.08041 &  1  \\
 \ion{Fe}{2} 1608$^a$ & 1608.4511 & 0.058 &  6 \\
 \ion{Fe}{2} 1611 & 1611.2005 & 1.36E-3 &  7  \\
   \ion{C}{1} 1656 & 1656.9283 & 0.1405 &  1  \\
 \ion{Al}{2} 1670$^a$ & 1670.7874 & 1.88 &  1 \\
 \ion{Ni}{2} 1703 & 1703.4050 & 0.006 &  4  \\
 \ion{Ni}{2} 1709 & 1709.6000 & 0.0324 &  4  \\
 \ion{Ni}{2} 1741 & 1741.5490 & 0.0427 &  4  \\
 \ion{Ni}{2} 1751 & 1751.9100 & 0.0277 &  4  \\
 \ion{Si}{2} 1808$^a$ & 1808.0126 & 2.186E-3 &  8 \\
 
 \ion{Mg}{1} 1827 & 1827.9351 & 2.420E-2 & 17 \\
 
  \ion{Si}{1} 1845 & 1845.5200 & 0.229 &  1  \\
 \ion{Al}{3} 1854 & 1854.7164 & 0.539 &  1  \\
 \ion{Al}{3} 1862 & 1862.7895 & 0.268 &  1  \\
 \ion{Fe}{2} 1901 & 1901.7730 & 1.009E-4 &  1  \\
 
 \ion{Ti}{2} 1910 & 1910.7800 & 0.2020 & 17 \\
 
 \ion{Zn}{2} 2026 & 2026.1360 & 0.489 &  9  \\
 \ion{Cr}{2} 2026 & 2026.2690 & 4.71E-3 & 10  \\
 \ion{Mg}{1} 2026 & 2026.4768 & 0.1120 & 1 \\
 \ion{Cr}{2} 2056 & 2056.2539 & 0.105 &  9  \\
 \ion{Cr}{2} 2062 & 2062.2340 & 0.078 &  9  \\
 \ion{Zn}{2} 2062 & 2062.6640 & 0.256 &  9  \\
 \ion{Cr}{2} 2066 & 2066.1610 & 0.0515 &  9  \\
 \ion{Fe}{2} 2249 & 2249.8768 & 1.821E-3 & 11  \\
 \ion{Fe}{2} 2260 & 2260.7805 & 2.44E-3 & 11  \\
 
 \ion{C}{2]} 2325 & 2325.4029 & 4.780E-8 & 17 \\
 \ion{C}{2]*} 2326 & 2326.1126 & 5.520E-8 & 17 \\
 \ion{C}{2]*} 2328 & 2328.8374 & 2.720E-8 & 17 \\
 \ion{Si}{2]} 2335 & 2335.1230 & 4.250E-6 & 17 \\
 
 \ion{Fe}{2} 2344$^a$ & 2344.2140 & 0.114 & 12 \\
 \ion{Fe}{2} 2374 & 2374.4612 & 0.0313 & 12  \\
 \ion{Fe}{2} 2382$^a$ & 2382.7650 & 0.32 & 12 \\
 \ion{Mn}{2} 2576 & 2576.8770 & 0.3508 &  1  \\
 \ion{Fe}{2} 2586$^a$ & 2586.6500 & 0.0691 & 12 \\
 \ion{Mn}{2} 2594 & 2594.4990 & 0.271 &  1  \\
 \ion{Fe}{2} 2600$^a$ & 2600.1729 & 0.239 & 12 \\
 \ion{Mn}{2} 2606 & 2606.4620 & 0.1927 &  1  \\
 \ion{Mg}{2} 2796$^a$ & 2796.3520 & 0.6123 & 13 \\
 \ion{Mg}{2} 2803$^a$ & 2803.5310 & 0.3054 & 13 \\
  \ion{Mg}{1} 2852 & 2852.9642 & 1.81 &  1  \\
 \ion{Ti}{2} 3073 & 3073.8770 & 0.1091 &  1  \\
 \ion{Ti}{2} 3230 & 3230.1310 & 0.0687 &  15  \\
 \ion{Ti}{2} 3242 & 3242.9290 & 0.232 &  16  \\
 \ion{Ti}{2} 3384 & 3384.7400 & 0.358 &  17  \\
\enddata
\tablenotetext{a}{: Member of the {\it{sdss\_metals}} search (see $\S$~3.1).}
\tablerefs{
1: \cite{morton91}; 
2: \cite{trp96}; 
3: \cite{fedchak99}; 
4: \cite{fedchak00}; 
5: \cite{schect98}; 
6: \cite{bergs96}; 
7: \cite{raassen98}; 
8: \cite{bergs93}; 
9: \cite{bergs93b}; 
10: \cite{verner94};
11: \cite{bergs94}; 
12: \cite{morton01};
13: \cite{verner96};
14: \cite{dessauges06};
15: \cite{bizzarri93};
16: \cite{pickering02};
17: \cite{morton03}}
\end{deluxetable}

\clearpage

\begin{deluxetable}{lccl}
\tablewidth{0pc}
\tablecaption{Summary of 
Subjective Metal Ratings of SDSS Candidates \label{mtl_rat}}
\tabletypesize{\footnotesize}
\tablehead{\colhead{Metals} &\colhead{Label} 
&\colhead{Rating} &\colhead{Description} 
}
\startdata
Bizarre & B & 0 & Noise, sky, or unclear detection \\
None    & N & 1 & No metals observed \\
Weak    & W & 2 & Usually weak \ion{Al}{2} 1670 and/or few other strong metals \\
Medium  & M & 3 & Usually strong \ion{Al}{2} 1670 or \ion{Fe}{2}, \ion{Mg}{2} but no significant \ion{Si}{2} 1808 \\
Strong  & S & 4 & Strong \ion{Si}{2} 1808 ($F_{min}/F_q \approx 0.90$), perhaps weak \ion{Zn}{2} 2026 \\
Very Strong & VS & 5 & Very strong \ion{Si}{2} 1808 ($F_{min}/F_q \le 0.85$) and likely \ion{Zn}{2} 2026 \\
\enddata
\end{deluxetable}


\begin{deluxetable}{cccccccccc}
\tablewidth{0pc}
\tablecaption{Metal-Strong Candidates from SDSS-DR3 \label{all_DR3}}
\tabletypesize{\footnotesize}
\tablehead{\colhead{SDSS plate} &\colhead{MJD} &\colhead{SDSS fiber} &\colhead{RA} &\colhead{Dec} &\colhead{$r$} &\colhead{$z_{em}$} &\colhead{$z_{abs}$} &\colhead{Quality$^a$} &\colhead{Metals$^b$} \\
&&& (J2000) & (J2000) & (mag) }
\startdata
 651 &52139 &494 &00:08:15.33 &$-09:58:54.0 $&18.38 &1.951 &1.768 &10 &5\\
 388 &51792 &607 &00:10:17.80 &$+01:04:50.7 $&18.84 &1.817 &1.687 &10 &5\\
 389 &51793 &332 &00:10:25.93 &$+00:54:47.6 $&19.09 &2.847 &2.154 & 9 &5\\
 752 &52247 &194 &00:13:41.74 &$+14:35:31.3 $&19.39 &1.933 &1.922 &10 &4\\
 389 &51793 &497 &00:15:49.08 &$+00:17:31.9 $&19.64 &3.066 &2.338 & 7 &5\\
 389 &51793 &178 &00:16:02.40 &$-00:12:24.9 $&18.03 &2.087 &1.973 &10 &5\\
 753 &52227 &430 &00:20:28.97 &$+15:34:35.9 $&18.79 &1.764 &1.652 &10 &5\\
 753 &52227 &550 &00:26:15.58 &$+15:27:13.5 $&19.85 &2.914 &1.954 & 9 &4\\
 418 &51813 &452 &00:37:49.19 &$+15:52:08.4 $&20.03 &4.072 &3.816 &17 &5\\
 655 &52160 & 51 &00:42:05.22 &$-10:39:57.5 $&19.17 &2.490 &2.283 &10 &5\\
 656 &52147 &269 &00:42:19.74 &$-10:20:09.4 $&18.66 &3.881 &2.753 &14 &4\\
 655 &52160 &635 &00:43:49.53 &$-09:37:44.0 $&18.49 &2.130 &1.764 &10 &5\\
 393 &51793 &495 &00:44:39.32 &$+00:18:22.7 $&18.20 &1.866 &1.725 &10 &4\\
 393 &51793 & 62 &00:47:15.88 &$-00:36:44.0 $&18.73 &2.198 &2.029 &10 &5\\
 395 &51783 &121 &00:57:09.50 &$-00:54:50.9 $&18.86 &1.885 &1.791 &10 &4\\
 420 &51869 & 12 &00:57:57.32 &$+14:19:00.0 $&19.38 &2.154 &1.782 & 9 &5\\
 395 &51783 &445 &00:58:14.31 &$+01:15:30.3 $&17.69 &2.495 &2.011 &10 &5\\
 658 &52143 &490 &00:59:45.10 &$-09:51:55.1 $&20.61 &3.036 &2.979 & 8 &5\\
 396 &51813 &535 &01:06:48.02 &$+00:46:27.9 $&18.59 &1.877 &1.774 &10 &5\\ 
\enddata
\tablenotetext{a}{The automatically-assigned overall quality from 
{\it{sdss\_search}}; 18.0 is the highest rating (for metal-strong systems showing 
corresponding candidate DLAs detected by {\it{sdss\_dla}}), 
otherwise 10.0 for metal-strong systems without \lya\ coverage.  Three quality points 
are deducted if no candidate DLA is detected when \lya\ coverage is present.}
\tablenotetext{b}{4=`strong'; 5=`very strong'; determined from 
visual inspection of our SDSS sample}
\tablecomments{The complete version of this table is in the 
electronic edition of 
the paper.  The printed edition contains only a sample.}
\end{deluxetable}

\clearpage

\begin{deluxetable}{ccccccccccc}
\tablecaption{Keck ESI Observations \label{tab:obs}}
\tablewidth{0pc}
\tablehead{\colhead{RA} &\colhead{Dec.} &\colhead{} &\colhead{SDSS} &\colhead{} &\colhead{$z_{em}$} 
&\colhead{$r$} &\colhead{Metal$^a$} &\colhead{Exp.} 
&\colhead{Slit width} &\colhead{Obs. date} \\ 
(2000) & (2000) & Plate & MJD & Fiber & & (Mag.) & Rating & (s) & $''$ & (UT)}
\startdata
00:08:15.33 & -09:58:54.3 & 651 & 52141 & 494 & 1.95 & 18.38 & 5 & 900 & 0.75 & Dec. 20, 2003 \\
00:16:02.40 & -00:12:25.0 & 389 & 51795 & 178 & 2.09 & 18.02 & 5 & 600 & 0.50 & Sept. 10, 2004 \\
00:20:28.96 & +15:34:35.8 & 753 & 52233 & 430 & 1.76 & 18.78 & 5 & 1800 & 0.50 & Sept. 10, 2004 \\
00:44:39.32 & +00:18:22.8 & 393 & 51794 & 495 & 1.87 & 18.20 & 4 & 600 & 0.75 & Dec. 20, 2003 \\
00:58:14.31 & +01:15:30.2 & 395 & 51783 & 445 & 2.49 & 17.66 & 5 & 900 & 0.50 & Dec. 20, 2003 \\
01:20:20.37 & +13:24:33.5 & 424 & 51893 & 286 & 2.57 & 19.23 & 5 & 900 & 0.75 & Dec. 20, 2003 \\
02:25:54.85 & +00:54:51.9 & 406 & 51869 & 572 & 2.97 & 18.92 & 4 & 1800 & 0.50 & Dec. 20, 2003 \\
03:16:09.83 & +00:40:43.1 & 413 & 51929 & 387 & 2.92 & 18.67 & 4 & 1200 & 0.50 & Dec. 20, 2003 \\
08:40:32.96 & +49:42:52.9 & 445 & 51873 & 043 & 2.08 & 19.00 & 5 & 1800 & 0.75 & Dec. 20, 2003 \\
08:44:07.29 & +51:53:11.2 & 447 & 51877 & 272 & 3.20 & 19.22 & 5 & 1800 & 0.75 & Dec. 20, 2003 \\
09:12:47.59 & -00:47:17.3 & 472 & 51955 & 210 & 2.86 & 18.67 & 5 & 900 & 0.50 & Dec. 20, 2003 \\
09:27:05.90 & +56:21:14.1 & 451 & 51908 & 071 & 2.28 & 18.24 & 5 & 900 & 0.75 & Dec. 20, 2003 \\
10:42:52.32 & +01:17:36.5 & 506 & 52022 & 137 & 2.44 & 18.42 & 5 & 600 & 0.50 & Dec. 20, 2003 \\
10:49:15.43 & -01:10:38.1 & 275 & 51910 & 006 & 2.12 & 17.78 & 5 & 600 & 0.50 & Dec. 20, 2003 \\
11:51:22.14 & +02:04:26.3 & 515 & 52051 & 223 & 2.40 & 18.60 & 4 & 600 & 0.50 & Dec. 20, 2003 \\
12:35:59.29 & +00:17:16.4 & 290 & 51941 & 463 & 2.26 & 18.84 & 4 & 600 & 0.50 & Dec. 20, 2003 \\
12:49:24.86 & -02:33:39.7 & 336 & 51999 & 073 & 2.12 & 17.79 & 5 & 900 & 0.50 & Dec. 20, 2003 \\
14:35:12.94 & +04:20:36.9 & 585 & 52027 & 628 & 1.95 & 19.04 & 5 & 600 & 0.50 & Dec. 20, 2003 \\
16:10:09.42 & +47:24:44.4 & 813 & 52354 & 621 & 3.22 & 18.75 & 5 & 1800 & 0.50 & Sept. 10, 2004 \\
16:17:17.83 & +00:28:27.2 & 346 & 51693 & 488 & 1.94 & 19.07 & 4 & 850 & 0.50 & Sept. 10, 2004 \\
16:58:16.47 & +34:28:09.8 & 972 & 52435 & 480 & 1.70 & 18.40 & 4 & 1200 & 0.50 & Sept. 11, 2004 \\
17:09:09.28 & +32:58:03.4 & 973 & 52426 & 629 & 1.89 & 19.22 & 5 & 1800 & 0.50 & Sept. 11, 2004 \\
20:44:31.12 & -05:42:39.7 & 634 & 52164 & 634 & 1.90 & 18.67 & 4 & 1200 & 0.50 & Sept. 10, 2004 \\
20:59:22.42 & -05:28:42.7 & 636 & 52176 & 610 & 2.54 & 19.12 & 5 & 1200 & 0.50 & Sept. 11, 2004 \\
21:00:25.03 & -06:41:45.9 & 637 & 52174 & 370 & 3.14 & 18.19 & 5 & 1200 & 0.50 & Sept. 10, 2004 \\
22:22:56.11 & -09:46:36.2 & 720 & 52206 & 203 & 2.93 & 17.97 & 4 & 600 & 0.50 & Sept. 10, 2004 \\
22:44:52.22 & +14:29:15.1 & 740 & 52263 & 390 & 1.96 & 18.92 & 4 & 1800 & 0.50 & Sept. 11, 2004 \\
\enddata
\tablenotetext{a}{See Table~\ref{mtl_rat}}
\end{deluxetable}


 
\clearpage
\begin{deluxetable}{lcccccccc}
\tablecaption{Zn Analysis\label{tab:znanly}}
\tablewidth{0pc}
\tablehead{\colhead{QSO} & 
\colhead{$W_{2026}$} &
\colhead{$W_{2852}$} & \colhead{$W({\rm Mg})^a$} & \colhead{$\log \N{Zn^+}_{2026}$} & \colhead{$W_{2062}$}
& \colhead{$W({\rm Cr})^b$} & \colhead{$\log \N{Zn^+}_{2062}$} & \colhead{$\log\N{Zn^+}$} \\ 
& (m\AA) & (m\AA) & (m\AA) & & (m\AA) & (m\AA) }
\startdata
FJ0812+32&$ 173\pm  9$&$...$&$...$&$ 12.99 \pm 0.02$&$  72\pm  6$&$ 69$&$ 11.46 \pm 1.12$&$ 12.84 \pm 0.03$\\
SDSS0008-0958&$ 383\pm 20$&$ 934\pm 27$&$ 37$&$<13.34$&$ 280\pm 22$&$159$&$ 13.10 \pm 0.09$&$ 13.10 \pm 0.09$\\
SDSS0016-0012&$ 190\pm 36$&$ 997\pm 25$&$ 39$&$ 12.93 \pm 0.10$&$   8\pm 34$&$...$&$<13.02$&$ 12.93 \pm 0.10$\\
SDSS0020+1534&$ 170\pm 36$&$ 843\pm 35$&$ 37$&$<13.06$&$...$&$ 88$&$...$&$<13.06$\\
SDSS0044+0018&$  50\pm 23$&$ 597\pm 28$&$ 21$&$<12.59$&$...$&$...$&$...$&$<12.59$\\
SDSS0058+0115&$  90\pm 12$&$ 675\pm 22$&$ 24$&$ 12.57 \pm 0.08$&$  93\pm 16$&$101$&$<12.90$&$ 12.57 \pm 0.08$\\
SDSS0120+1324&$  65\pm 59$&$ 616\pm 61$&$ 24$&$<13.00$&$  58\pm 55$&$...$&$<13.23$&$<13.00$\\
SDSS0225+0054&$ 125\pm 30$&$...$&$...$&$ 12.85 \pm 0.10$&$...$&$178$&$...$&$ 12.85 \pm 0.10$\\
SDSS0316+0040&$  48\pm 29$&$ 691\pm 40$&$ 32$&$<12.69$&$   1\pm 30$&$...$&$<12.97$&$<12.69$\\
SDSS0840+4942&$ 224\pm 74$&$1091\pm 70$&$ 46$&$<13.26$&$  88\pm 59$&$...$&$<13.26$&$<13.26$\\
SDSS0844+5153&$...$&$...$&$...$&$...$&$ 187\pm 22$&$184$&$<13.08$&$<13.08$\\
SDSS0912-0047&$ 168\pm 27$&$ 543\pm 48$&$< 23$&$ 12.91 \pm 0.09$&$ 130\pm 34$&$...$&$ 13.13 \pm 0.11$&$ 12.96 \pm 0.07$\\
SDSS0927+5621&$  94\pm 28$&$ 674\pm 47$&$ 25$&$<12.68$&$ 138\pm 29$&$ 78$&$<13.18$&$<12.68$\\
SDSS1042+0117&$  72\pm 31$&$ 716\pm101$&$ 34$&$<12.72$&$  83\pm 28$&$ 88$&$<13.13$&$<12.72$\\
SDSS1049-0110&$ 132\pm 52$&$ 754\pm 28$&$ 28$&$<12.94$&$ 118\pm 37$&$128$&$<13.35$&$<12.94$\\
SDSS1151+0204&$ 375\pm 56$&$1606\pm 57$&$ 66$&$<13.37$&$ 333\pm 49$&$161$&$<13.41$&$<13.37$\\
SDSS1235+0017&$...$&$ 188\pm 66$&$<  7$&$...$&$  46\pm 33$&$...$&$<13.01$&$<13.01$\\
SDSS1249-0233&$ 208\pm 19$&$ 695\pm 24$&$ 28$&$ 13.01 \pm 0.04$&$ 279\pm 19$&$238$&$<12.97$&$ 13.01 \pm 0.04$\\
SDSS1435+0420&$ 286\pm 97$&$ 948\pm 51$&$ 41$&$<13.21$&$ 272\pm 57$&$155$&$<13.48$&$<13.21$\\
SDSS1610+4724&$ 470\pm 27$&$ 361\pm143$&$ 19$&$ 13.40 \pm 0.03$&$...$&$218$&$...$&$ 13.40 \pm 0.03$\\
SDSS1617+0028&$ 269\pm 36$&$1332\pm 41$&$ 58$&$<13.20$&$ 169\pm 51$&$124$&$<13.42$&$<13.20$\\
SDSS1658+3428&$ 378\pm 35$&$1252\pm 25$&$ 50$&$<13.35$&$  50\pm 26$&$...$&$<12.92$&$<12.92$\\
SDSS1709+3258&$ 309\pm 22$&$ 598\pm 30$&$ 24$&$ 13.21 \pm 0.03$&$ 324\pm 24$&$215$&$ 13.05 \pm 0.12$&$ 13.19 \pm 0.03$\\
SDSS2044-0542&$ 113\pm 22$&$ 333\pm 29$&$ 13$&$ 12.75 \pm 0.09$&$ 114\pm 24$&$104$&$<13.11$&$ 12.75 \pm 0.09$\\
SDSS2059-0529&$ 138\pm 37$&$ 600\pm 60$&$ 25$&$ 12.81 \pm 0.14$&$  60\pm 35$&$104$&$<13.04$&$ 12.81 \pm 0.14$\\
SDSS2100-0641&$ 213\pm 16$&$...$&$...$&$<13.14$&$...$&$145$&$...$&$<13.14$\\
SDSS2222-0946&$ 103\pm 17$&$ 787\pm 63$&$ 33$&$<12.77$&$...$&$...$&$...$&$<12.77$\\
SDSS2244+1429&$ 261\pm 18$&$ 759\pm 32$&$ 32$&$ 13.11 \pm 0.03$&$ 114\pm 22$&$ 78$&$<13.07$&$ 13.11 \pm 0.03$\\
\enddata
\tablenotetext{a}{Estimated equivalent width for the MgI~2026 transition from the column density measured from
 MgI~2852.  For those cases where we expect a saturation corection, we have incremented the column density by 0.1~dex.}
\tablenotetext{b}{Estimated equivalent width for the CrII~2062 transition from the measured Cr$^+$ column densities.} 
\tablecomments{This analysis assumes the linear curve-of-growth
 for all equivalent width calculations.  We may be underestimating 
the Zn$^+$ column density of those systems with equivalent width $>300$ m\AA.
Entries with `...' are cases of upper limits to the value (i.e. non-detections).
}
\end{deluxetable}

\clearpage

\begin{deluxetable}{lccccccccccc}
\tablecaption{\ion{Si}{2} 1808 and \ion{Zn}{2} 2026 as Metal-Strong Indicators \label{tab:ind}}
\tablecolumns{13}
\tablewidth{555pt}
\tabletypesize{\scriptsize}
\tablehead{\colhead{QSO name} &\colhead{SDSS} &\colhead{$z_{em}$}
&\colhead{$z_{abs}$}
&\colhead{$r$}
&\colhead{log$\mnhi$}
&\colhead{log$\N{Si II\ 1808}$}
&\colhead{$W_{\rm S}(\rm 1808)$}
&\colhead{$W_{\rm E}(\rm 1808)$}
&\colhead{log$\N{Zn II\ 2026}$}
&\colhead{$W_{\rm S}(\rm 2026)$}
&\colhead{$W_{\rm E}(\rm 2026)$}}
\startdata
FJ0812+32 & 861-333 &  2.70 & 2.626 & 17.46 & 21.35* &$ >15.78 $&$   250\pm 50 $&$  285\pm  8$&$13.04 \pm  0.02 $&$   190\pm 50 $&$  173\pm  9$\\
SDSS0008-0958 & 651-494 &  1.95 & 1.768 & 18.38 & ... &$ 16.08 \pm  0.03 $&$   540\pm 50 $&$  603\pm 33$&$<13.34 $&$   300\pm 50 $&$  383\pm 20$\\
SDSS0016-0012 & 389-178 &  2.09 & 1.970 & 18.02 & 20.83** &$ 15.48 \pm  0.11 $&$   330\pm 50 $&$  165\pm 46$&$12.93 \pm  0.10 $&$   340\pm 50 $&$  190\pm 36$\\
SDSS0020+1534 & 753-430 &  1.76 & 1.652 & 18.78 & ... &$ 15.35 \pm  0.09 $&$   360\pm 50 $&$  119\pm 26$&$<13.06 $&$   230\pm 50 $&$  170\pm 36$\\
SDSS0044+0018 & 393-495 &  1.87 & 1.725 & 18.20 & ... &$ <15.51 $&$   100\pm 50 $&$  183\pm 31$&$<12.59 $&$    60\pm 50 $&$   50\pm 23$\\
SDSS0058+0115 & 395-445 &  2.49 & 2.011 & 17.66 & ... &$ <15.82 $&$   370\pm 50 $&$  361\pm 25$&$12.57 \pm  0.08 $&$   110\pm 50 $&$   90\pm 12$\\
SDSS0120+1324 & 424-286 &  2.57 & 2.000 & 19.23 & ... &$ <16.04 $&$   250\pm 50 $&$  354\pm 68$&$<13.00 $&$   200\pm 50 $&$   65\pm 59$\\
SDSS0225+0054 & 406-572 &  2.97 & 2.714 & 18.92 & 21.00*** &$ 15.61 \pm  0.07 $&$   260\pm 50 $&$  227\pm 35$&$12.85 \pm  0.10 $&$   240\pm 50 $&$  125\pm 30$\\
SDSS0316+0040 & 413-387 &  2.92 & 2.181 & 18.67 & ... &$ 15.41 \pm  0.08 $&$   130\pm 50 $&$  137\pm 25$&$<12.69 $&$   150\pm 50 $&$   48\pm 29$\\
SDSS0840+4942 & 445-043 &  2.08 & 1.851 & 19.00 & ... &$ 15.83 \pm  0.07 $&$   250\pm 50 $&$  319\pm 57$&$<13.26 $&$   200\pm 50 $&$  224\pm 74$\\
SDSS0844+5153 & 447-272 &  3.21 & 2.775 & 19.22 & 21.45*** &$ 15.97 \pm  0.02 $&$   370\pm 50 $&$  465\pm 19$& sky & sky & sky \\
SDSS0912-0047 & 472-210 &  2.86 & 2.071 & 18.67 & ... &$ >15.80 $&$   430\pm 50 $&$  285\pm 33$&$12.91 \pm  0.09 $&$   150\pm 50 $&$  168\pm 27$\\
SDSS0927+5621 & 451-071 &  2.28 & 1.775 & 18.24 & 19.00* &$ <15.20 $&$   270\pm 50 $&$   70\pm 33$&$<12.68 $&$   120\pm 50 $&$   94\pm 28$\\
SDSS1042+0117 & 506-137 &  2.44 & 2.267 & 18.42 & 20.75*** &$ 15.47 \pm  0.09 $&$   230\pm 50 $&$  159\pm 33$&$<12.72 $&$   100\pm 50 $&$   72\pm 31$\\
SDSS1049-0110 & 275-006 &  2.12 & 1.658 & 17.78 & ... &$ 15.77 \pm  0.03 $&$   290\pm 50 $&$  330\pm 26$&$<12.94 $&$   190\pm 50 $&$  132\pm 52$\\
SDSS1151+0204 & 515-223 &  2.40 & 1.968 & 18.60 & ... &$ 15.83 \pm  0.08 $&$   400\pm 50 $&$  350\pm 72$&$<13.37 $&$   330\pm 50 $&$  375\pm 56$\\
SDSS1235+0017 & 290-463 &  2.27 & 2.023 & 18.84 & ... &$ <15.38 $&$   280\pm 50 $&$   43\pm 48$& blend & blend & blend \\
SDSS1249-0233 & 336-073 &  2.12 & 1.781 & 17.79 & 21.45* &$ 15.80 \pm  0.03 $&$   400\pm 50 $&$  313\pm 23$&$13.01 \pm  0.04 $&$   210\pm 50 $&$  208\pm 19$\\
SDSS1435+0420 & 585-628 &  1.95 & 1.656 & 19.04 & 21.25* &$ 15.92 \pm  0.07 $&$   510\pm 50 $&$  408\pm 54$&$<13.21 $&$   330\pm 50 $&$  286\pm 97$\\
SDSS1610+4724 & 813-621 &  3.22 & 2.508 & 19.22 & 21.15 &$ >16.15 $&$   630\pm 50 $&$  632\pm 23$&$13.40 \pm  0.03 $&$   520\pm 50 $&$  470\pm 27$\\
SDSS1617+0028 & 346-488 &  1.94 & 1.616 & 19.07 & ... &$ 15.89 \pm  0.04 $&$   450\pm 50 $&$  385\pm 36$&$<13.20 $&$   240\pm 50 $&$  269\pm 36$\\
SDSS1658+3428 & 972-480 &  1.70 & 1.658 & 18.40 & ... &$ 15.79 \pm  0.03 $&$   360\pm 50 $&$  338\pm 25$&$<13.35 $&$   280\pm 50 $&$  378\pm 35$\\
SDSS1709+3258 & 973-629 &  1.89 & 1.830 & 19.22 & ... &$ >16.11 $&$   550\pm 50 $&$  528\pm 20$&$13.21 \pm  0.03 $&$   370\pm 50 $&$  309\pm 22$\\
SDSS2044-0542 & 634-634 &  1.91 & 1.787 & 18.81 & ... &$ 15.68 \pm  0.05 $&$   360\pm 50 $&$  217\pm 26$&$12.75 \pm  0.09 $&$   300\pm 50 $&$  113\pm 22$\\
SDSS2059-0529 & 636-610 &  2.54 & 2.210 & 19.01 & 20.80 &$ 15.36 \pm  0.10 $&$   370\pm 50 $&$  113\pm 31$&$12.81 \pm  0.14 $&$   280\pm 50 $&$  138\pm 37$\\
SDSS2100-0641 & 637-370 &  3.14 & 3.092 & 18.12 & 21.05 &$ 15.89 \pm  0.02 $&$   420\pm 50 $&$  422\pm 15$&$<13.14 $&$   300\pm 50 $&$  213\pm 16$\\
SDSS2222-0946 & 720-203 &  2.93 & 2.354 & 17.97 & 20.50 &$ 15.45 \pm  0.05 $&$   260\pm 50 $&$  156\pm 19$&$<12.77 $&$   140\pm 50 $&$  103\pm 17$\\
SDSS2244+1429 & 740-390 &  1.96 & 1.816 & 18.92 & ... &$ 15.62 \pm  0.04 $&$   260\pm 50 $&$  225\pm 20$&$13.11 \pm  0.03 $&$   220\pm 50 $&$  261\pm 18$\\
\enddata
\tablecomments{Only confirmed log$\mnhi$ values are shown; 
all have an error of $\pm 0.15$ dex except SDSS0016--0012 ($\pm 0.05$ dex), SDSS0927+5621
($^{+0.10}_{-0.25}$ dex), 
\\and SDSS2059--0529 ($\pm 0.20$ dex).
All metal-line abundances can be found in the electronic edition of the paper.  
$W_r$ values are given in m\AA\ and do not 
\\account for minor blends, 
low SNR or saturation effects.
This table does not account for possible $W_r$(MgI 2026) contribution to $W_r$(ZnII 2026) (although
\\log$\N{Zn II\ 2026}$ values 
{\it{are}} blend-corrected);
this issue 
\\is dealt with in Table~\ref{tab:znanly}.
\vspace{0.3cm}}
\tablenotetext{*}{Measurement from Prochaska et al.\ (2006)}
\tablenotetext{**}{Measurement from Petitjean et al. (2002)}
\tablenotetext{***}{Measurement from Prochaska \& Herbert-Fort (2004)}
\end{deluxetable}

\end{document}